\documentclass[rmp,twocolumn,showpacs,preprintnumbers,amsmath,amssymb,floatfix]{revtex4-2}

\usepackage{graphicx}
\usepackage{dcolumn}
\usepackage{bm}
\usepackage{color}
\usepackage[dvipsnames]{xcolor}
\usepackage{hyperref}
\hypersetup{colorlinks=true,urlcolor=blue,citecolor=blue}
\usepackage[colorinlistoftodos]{todonotes}
\usepackage{soul}

\begin{document}

\preprint{DRAFT}

\title{Colloquium: Quantum Properties and Functionalities of Magnetic Skyrmions} 


\author{Alexander P. Petrovi\'c$^{1}$}
\affiliation{Division of Physics and Applied Physics, School of Physical and Mathematical Sciences,
Nanyang Technological University, 637371 Singapore 
\\and Department of Physics and Astronomy, University of Wyoming, WY 82071, USA
\\and Center for Quantum Information Science and Engineering, University of Wyoming, WY 82071, USA} 

\author{Christina Psaroudaki$^{1}$}
\affiliation{Laboratoire de Physique de l’\'{E}cole Normale Sup\'{e}rieure, ENS, Universit\'{e} PSL, CNRS, Sorbonne Universit\'{e}, Universit\'{e} de Paris, F-75005 Paris, France}
\footnote[1]{These authors contributed equally to this work}

\author{Peter Fischer}
\affiliation{Materials Sciences Division, Lawrence Berkeley National Laboratory, Berkeley CA 94720, \\ USA 
\\and Physics Department, University of California Santa Cruz, Santa Cruz CA 95064, USA}

\author{Markus Garst}
\affiliation{Institute of Theoretical Solid State Physics\\ and Institute for Quantum Materials and Technology, Karlsruhe Institute of Technology, D-76131 Karlsruhe, Germany}
\email{markus.garst@kit.edu}

\author{Christos Panagopoulos}
\affiliation{Division of Physics and Applied Physics, School of Physical and Mathematical Sciences,
Nanyang Technological University, 637371 Singapore}
\email{christos@ntu.edu.sg}

\date{\today}

\begin{abstract}
Competing magnetic interactions may stabilize smooth magnetization textures that can be characterized by a topological winding number. Such textures, which are spatially localized within a two-dimensional plane, are commonly known as skyrmions. On the classical level, their significance for fundamental science and their potential for applications, ranging from spintronic devices to unconventional computation platforms, have been intensively investigated in recent years. This Colloquium considers quantum effects associated with skyrmion textures: their theoretical origins, the experimental and material challenges associated with their detection, and the promise of exploiting them for quantum operations. Starting with classical skyrmions, we discuss their magnon and electron excitations and show how hybrid architectures offer new platforms for engineering quantum orders, including topological superconductivity. We then focus on the quantization of the skyrmion texture itself and formulate the long-time skyrmion dynamics in terms of collective coordinates. Next, we discuss the quantization of helicity and phenomena of macroscopic quantum tunneling: key concepts which fundamentally distinguish quantum skyrmions from their classical counterparts.  Looking ahead, we propose material classes suitable for the realization of skyrmions in quantum spin systems and identify device architectures with the promise of achieving quantum operations. We close by addressing the advances in experimental methods which will be prerequisite for resolving the quantum aspects of topological spin patterns, sensing their local dynamical response, and achieving their predicted functionalities in magnetic systems. 
\end{abstract}

\renewcommand{\vec}{\mathbf}

\maketitle 

\tableofcontents
\section{Introduction} 
Magnetic systems offer a fertile ground for investigating phenomena occurring at widely varying energy and length scales. These can range from interactions at the atomic level to processes observed on mesoscopic and macroscopic scales \cite{skjeltorp1998dynamical}. The ability to precisely design and synthesize magnetic nanostructures promises to unlock functionalities underpinned by quantum effects. It transpires that magnetic systems can often be quantized at different levels, as long as the energy involved does not surpass a certain threshold and the underlying microscopic structure of magnetic matter remains apparent \cite{chudnovsky_tejada_1998}. 

Magnetic skyrmions are among the smallest possible magnetic configurations in nature and therefore prime candidates to show coherent quantum dynamics involving a large ensemble of spins. These localized nanoscale textures are protected from annihilation by sizable energy barriers, which originate from the integer topological charge describing a quantized winding of the magnetization field \cite{doi:10.1098/rspa.1961.0018,Bogdanov1989a,Rossler2006}. Classical properties of skyrmions have been extensively discussed over the last decade and demonstrated in a plethora of systems, primarily motivated by the desire to develop particle-like information carriers in spintronic technologies \cite{Nagaosa2013,Wiesendanger2016,Fert2017,10.1063/1.5048972,Bogdanov2020,10.1063/5.0072735,finocchio2021magnetic}. Their typical sizes range from a few to tens of nanometers in diameter, depending on the host material parameters and external magnetic field~\cite{Wang2018}. 

The experimental observation of stable atomic-scale skyrmions in the low-temperature regime \cite{Kurumaji2019,Hirschberger2019} offers advantages for ultra-dense memories and poses new challenges for the theory of magnetism. At such small length scales and low temperatures, the often-overlooked quantum fluctuations can be significant, rendering classical micromagnetism inadequate. Quantum mechanics reveals itself in a wide range of phenomena in studies of magnetic skyrmions and is reflected in magnon properties~\cite{Diaz2020,Weber2022}, topological texture dynamics~\cite{PhysRevX.7.041045}, and the realization of controllable topological and quantum phase transitions \cite{PhysRevResearch.4.L032025}. 

Skyrmion textures of \emph{all} sizes are expected to remain robust over long lifetimes thanks to their large free energy barriers~\cite{Je2020,Sotnikov2023}, even though the continuous field concept of topological protection breaks down for nanoskyrmions comprising a finite number of discrete spins.  Quantum effects may also arise for systems with small spin quantum numbers, where classical vectors cannot replace spin operators. In this regime, the quantum nature of local spins leads to the formation of the quantum analog of the classical skyrmionic structure \cite{PhysRevB.103.L060404,PhysRevX.9.041063} with novel features including a rich spectrum of new non-classical states, superposed quantum states, and quantum phase transitions. 

Beyond such academic motivations, quantum effects can also deliver new technological applications for skyrmions, especially in information processing. Although quantum mechanics imposes the ultimate limitations to minimizing classical magnetic memory elements, it also introduces entirely new paradigms for topological \cite{PhysRevLett.117.017001,Rex2019,Petrovic2021a} or gate-based quantum computers \cite{PhysRevLett.127.067201,10.1063/5.0177864} built from skyrmions. We are hence equipped with a unique platform for exploring novel quantum phenomena in mesoscopic spin systems. In the quantum limit, magnetic skyrmions not only challenge established notions of topology, but also offer remarkable opportunities to merge the fields of spintronics, quantum computing, and strongly correlated systems.  For example, combining new skyrmionic logic components with quantum state memory elements (or memelements) may enable quantum advantage by storing and processing quantum information within the same device,  while interactions between magnetic skyrmions and topological solitons in superconductors can trigger the emergence of states exhibiting strongly-renormalized flux dynamics as well as non-Abelian anyonic quasiparticles. 

Delivering on this potential requires sustained development of platforms for creating, manipulating, and tuning skyrmions. The evolution of ultra-sensitive quantum sensors \cite{RevModPhys.89.035002, Rugar2004} and the advent of fully coherent x-ray free electron lasers \cite{AssefaRSI2022} now enable the detection of weak magnetic signals and local spin fluctuations \cite{SeabergPRL2017}, as well as the control and manipulation of spin ensembles at the nanoscale \cite{2401.04793}. In parallel, advances in material engineering of tailored magnetic systems - including many new skyrmion-hosting materials \cite{Tokura2021} - are now rendering experimental investigations of the quantum aspects of magnetic skyrmions feasible. Beyond observational studies of quantum coherence effects, hybrid skyrmion-based quantum architectures combining two or more physical systems \cite{PRXQuantum.3.040321, Garnier2019} offer new platforms for non-local quantum information storage with high resilience to decoherence. The electrical control over the helicity degree of freedom in quantum skyrmions opens another exciting new pathway toward the realization of quantum logic hardware. Excursions beyond two dimensions to create skyrmion strings, vortex loops, and hopfions \cite{Kent2021,Zheng2023} promise to further broaden the horizon for investigating novel quantum spin states. 

This Colloquium aims to offer a comprehensive overview of the quantum properties of magnetic skyrmions,  including their anticipated experimental observables and derived quantum devices. After a brief introduction to the stabilization mechanisms for classical skyrmions, we introduce architectures based on skyrmions coupled to quantum states, notably superconductor-skyrmion hybrids which allow the implementation of topological quantum computing. Moving on to a discussion of the semiclassical quantization of skyrmions and the potential for macroscopic quantum tunneling, we then unveil the prospects for realizing skyrmion qubits for quantum computing. Subsequently, we delve into the quantum nature of skyrmions comprising discrete spins with small spin quantum numbers. We additionally describe measurement techniques capable of exploring the quantum properties of topological spin textures, highlighting the synergy between theory and experiment. Finally, we outline several key scientific priorities whose accomplishment will allow quantum skyrmionics to deliver on its rich promise.

\section{Approaching the Quantum Limit with Classical Skyrmions}

Quantum phenomena involving magnetic skyrmions can emerge with different levels of complexity. In this section, we will first discuss the characterization of skyrmions on the classical level, as this forms the basis for a semiclassical description of various quantum effects. These effects become more pronounced for small skyrmions which are stable down to the lowest temperatures. For this reason, we will also shortly review common stabilization mechanisms favoring skyrmions with a small radius $r_{\rm sk}$, with a focus on thermodynamic skyrmion phases that persist down to zero temperature, $T=0$.

\subsection{Characterization of classical magnetic skyrmions}

\begin{figure*}[htbp]
\includegraphics[clip=true, width=2\columnwidth]{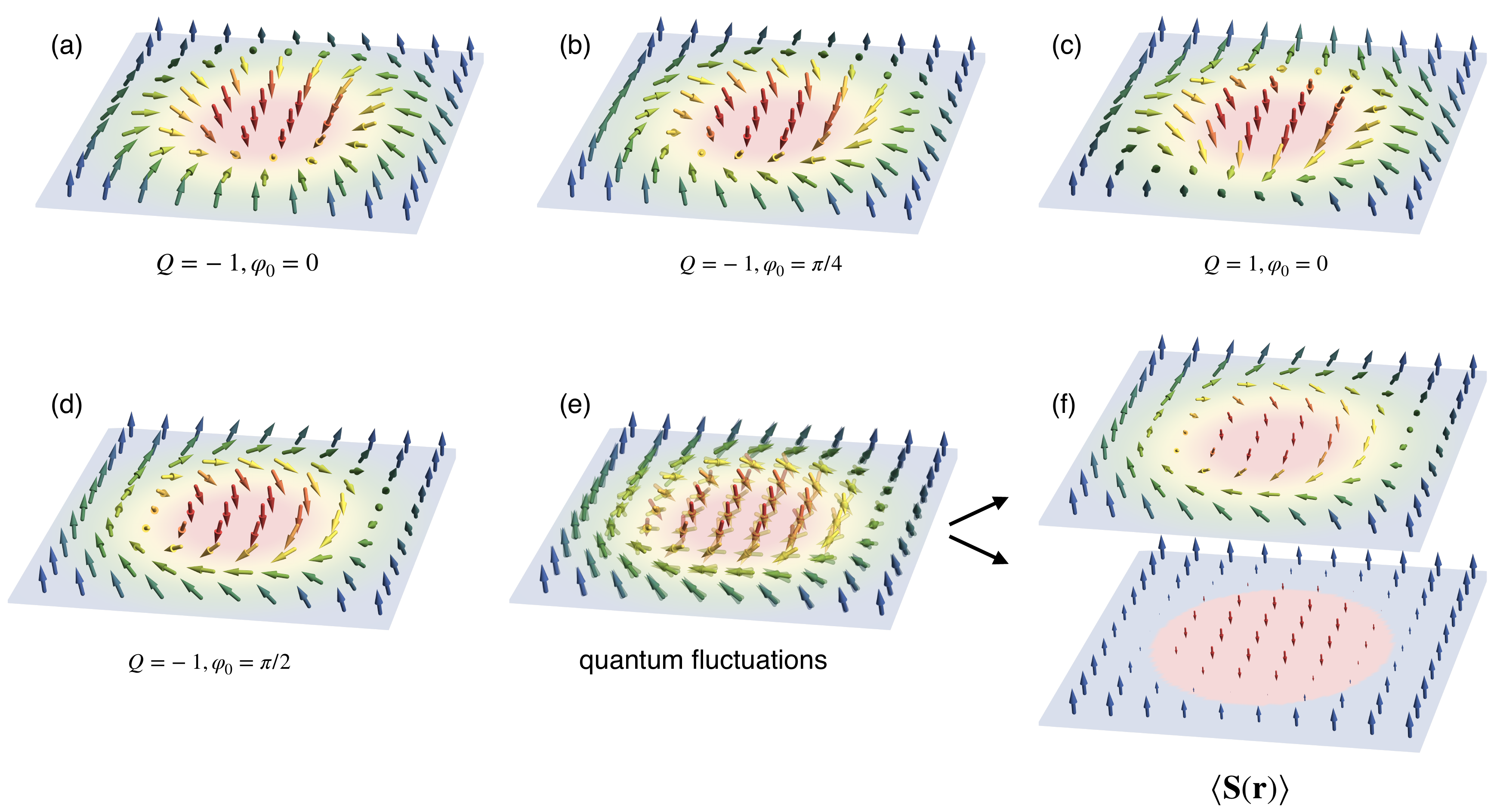}
\caption{A compendium of classical and quantum skyrmions. (a)-(d) Classical skyrmions with a local spin vector of fixed size for various values of the topological winding number $Q$ and helicity $\varphi_0$; (a) N\'eel skyrmion, (b) hybrid skyrmion, (c) antiskyrmion, (d) Bloch skyrmion. (e) Quantum fluctuations renormalize the texture leading to (f) expectation values of the local spin vector operator $\langle \hat{\vec S}(\vec r)\rangle$ with varying size. Depending on the stabilizing interactions,  $\langle \hat{\vec S}(\vec r)\rangle$ of a quantum skyrmion could differ substantially from the classical configuration: DMI stabilized (upper panel) and frustration stabilized (lower panel). The skyrmionic content of a quantum wavefunction is instead revealed by expectation values involving three spin operators on adjacent sites. 
} \label{Fig:Compendium}
\end{figure*}

Classically, magnetic textures are described by spatially varying magnetization profiles defined by the three-component unit vector field $\vec m(\vec r)$, which for each spatial position $\vec r$ is an element of the 2-sphere $S^2$, i.e., the order parameter manifold of a Heisenberg magnet. In two spatial dimensions, $d=2$, such textures can be topologically non-trivial because of the homotopy group $\Pi_2(S^2) = \mathbb{Z}$, where $\mathbb{Z}$ is the group of integers. For a profile $\vec m(\vec r) \to \vec m_0$ that approaches a specific orientation $\vec m_0$, e.g., $\vec m_0 = \hat z$, sufficiently fast for $|\vec r| \to \infty$, there exists a winding number or, equivalently, a topological charge  
\begin{equation}
Q= \int d^2\vec{r} \rho_{\rm top}(\vec{r}) , 
\label{TopCharge}
\end{equation}
that is an integer $Q \in \mathbb{Z}$. This winding number is given by a spatial integral over the topological charge density 
\begin{equation}
\rho_{\rm top} = \frac{1}{8\pi} \epsilon_{ij} \mathbf{m} \cdot (\partial_i \mathbf{m} \times \partial_j \mathbf{m} ) \,,
\label{TopChargeDensity}
\end{equation}
which is proportional to the solid angle spanned by $\vec m(\vec{r})$ in the vicinity of the real space position $\vec{r}$. 

Single classical skyrmions can be further characterized by three parameters: polarity $p$, vorticity $m$, and helicity $\varphi_0$~\cite{Nagaosa2013,Gobel2021}. These parameters are easily defined for skyrmions with circular symmetry. The unit vector field can be parametrized, $\vec m^T = (\sin\Theta \cos\Phi,\sin\Theta \sin\Phi,\cos\Theta)$, by azimuthal and polar angles, $\Phi$ and $\Theta$, respectively. 
In the case of a circular symmetric skyrmion with its center positioned at the origin, these angles might only depend on one of the polar coordinates, $\Phi = \Phi(\phi)$ and  $\Theta = \Theta(r)$ with the parametrization of the real space vector $\vec r^T = (r\cos\phi,r\sin\phi)$.
 If the magnetizations at the origin ($r=0$) and infinity are parallel to the $z$-axis, and the polar angle $\Theta$ covers its full range $(0,\pi)$ only once as a function of the radial variable $r$, the {\it polarity}
$p=-(1/2)\left[\cos\Theta(r)\right]^\infty_{r=0} = \hat z \vec m(0) = \pm 1$. $p$ hence characterizes whether the magnetization at the origin is aligned or anti-aligned with the $z$-axis. The {\it vorticity} $m=(1/2\pi)\left[\Phi(\phi)\right]^{2\pi}_{\phi=0} = 0, \pm1, \pm2, ...$ describes the number of complete rotations $\mathrm{\mathbf{m(r)}}$ performs in the horizontal plane while circling the skyrmion core at radial distance $r_{\rm sk}$, where $\Theta(r_{\rm sk})=\pi/2$. 
The quantity $r_{\rm sk}$ serves as a definition of the skyrmion radius. Since $\mathrm{\mathbf{m(r)}}$ is a continuous function, the vorticity must be an integer. The skyrmion winding number can then be expressed in terms of polarity and vorticity $Q = m p$ (Fig.~\ref{Fig:Compendium}).
For a polarity $p=-1$, the terminology {\it skyrmion} in a stricter sense is often associated with a magnetic texture with $m=1$ and charge $Q=-1$, whereas an antiskyrmion is a texture with vorticity $m=-1$ and charge $Q=1$. 
Finally, the {\it helicity} $\varphi_0$ is a phase factor that controls the azimuthal angle of $\mathrm{\mathbf{m(r)}}$:
\begin{equation}
\Phi(\phi) = m\phi + \varphi_0
\end{equation}
where $-\pi<\varphi_0\le\pi$.  Essentially, $\varphi_0$ describes the azimuthal angle between $\mathrm{\mathbf{m(r)}}$ and the radial vector $\mathrm{\mathbf{r}}$. Unlike $p$ and $m$, $\varphi_0$ can vary continuously: as we will shortly discover, helicity plays a crucial role in the semiclassical dynamics and functionality of skyrmions in certain magnetic systems with frustrated interactions.

In quantum mechanics, the local magnetic moment is instead given in terms of a spin vector operator $\hat{\vec S}(\vec r)$. The size of its local expectation value $\langle \hat {\vec S}(\vec r)\rangle$ is typically reduced from its maximal value $\hbar S$ by quantum fluctuations, with the size $S$ of the spin. These quantum fluctuations can lead to substantial renormalizations of $\langle \hat{\vec S}(\vec r)\rangle$ when compared with the classical skyrmion configurations (see Fig.~\ref{Fig:Compendium}(e) and (f)). In later sections, we will discuss whether and how the notion of a topological texture and the concept of a winding number $Q$ can be generalized to quantum skyrmions.

\subsection{Stabilization of skyrmions and low-$T$ phase diagrams}

Skyrmion stability hinges on a delicate balance between competing energy terms in magnetic materials. Depending on the crystal symmetry, skyrmions can be generated through several different mechanisms outlined below, notably via the competition of magnetic exchange and Dzyaloshinskii-Moriya interactions (DMI), or via the competition of various symmetric exchange interactions due to frustration and/or Fermi surface nesting. 

\begin{figure}[htbp]
\includegraphics[clip=true, width=1\columnwidth]{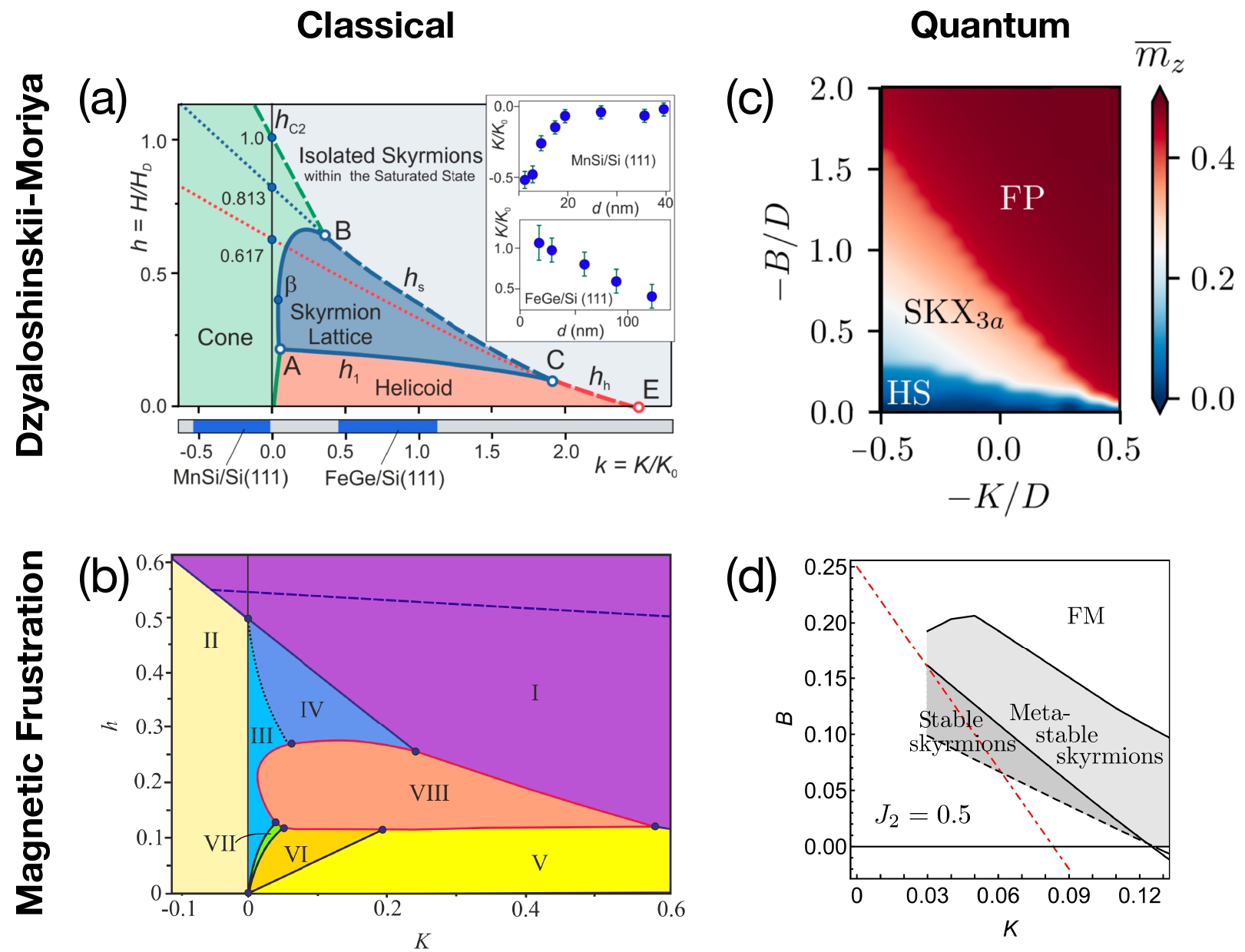}
\caption{\label{Fig:PhaseDiagrams} Skyrmion phase diagrams in the classical and quantum limits. Classical phase diagrams of (a) a magnet with DMI coupling and (b) a frustrated triangular magnet as a function of the magnetic field $h$ and uniaxial anisotropy $K$. The various phases are indicated by colored zones: the skyrmion lattice is denoted by blue shading in (a) and orange (labeled VIII) in (b). The corresponding phase diagrams for quantum spins with $S=1/2$ (c) with DMI of amplitude $D$ and (d) of the XXZ frustrated Heisenberg model, as a function of the magnetic field $B$ and anisotropy $K$ for finite systems sizes. Quantum skyrmions are stabilized as ground state lattice phases [pocket labeled $\mbox{SKX}_{3a}$ in (c) and dark shaded region in (d)], or as metastable excitations above the ferromagnetic background. Diagrams adapted from (a) \cite{PhysRevB.89.094411} (b) \cite{Leonov2015} (c) \cite{Haller2022} and (d) \cite{PhysRevX.9.041063}.}
\end{figure}

\subsubsection{Skyrmions in noncentrosymmetric crystals and interfaces}

In noncentrosymmetric crystals and at interfaces, the combined effect of spin-orbit coupling (SOC) and a lack of inversion symmetry generates a DMI of the form $\mathbf{D}_{ij} \cdot (\mathbf{S}_i \times \mathbf{S}_j$). In the continuum limit, this contributes terms in the free energy which are linear in the first-order spatial gradients of the magnetization, known as Lifshitz invariants~\cite{Dzyaloshinskii1964,Dzyaloshinskii1965}. Such terms favor a spatial spin rotation, which competes with the symmetric ferromagnetic exchange $J$ that in turn penalizes spatial variations of the magnetic moments. This competition can lead to the stabilization and formation of magnetic skyrmions, typically with a size of the order of $r_{\rm sk} \sim J/D$, i.e. it scales inversely with the size of the SOC. Increasing SOC is, thus, a common strategy to decrease the size of the resulting magnetic skyrmions. 
The crystal symmetry governs the profile and orientation of $\mathbf{D}$ in a given material resulting in distinct Lifshitz invariants \cite{Bogdanov1989a} which favor skyrmionic textures with, importantly, a specific value of the helicity. Crystals with chiral point groups $T$, $O$, or $D_n$ favor Bloch skyrmions with helicity $\varphi_0 = \pm \pi/2$, whereas crystals with polar point groups $C_{nv}$ instead favor N\'eel skyrmions with helicity $\varphi_0 = 0, \pi$, see Fig.~\ref{Fig:Compendium}(a). The point group $C_n$, which is both polar and chiral, prefers skyrmions with a fixed but non-universal value for the helicity $\varphi_0$. Finally, the DMI allowed for the point groups $D_{2d}$ and $S_4$ stabilize antiskyrmions, with helicity $\varphi_0 = \pm \pi/2$ for the former and a non-universal helicity for the latter. 

 Important examples of cubic materials with tetrahedral symmetry $T$ are the B20 compounds, e.g., MnSi and FeGe, as well as the magnetic insulator Cu$_2$OSeO$_3$. These materials share a similar magnetic phase diagram. A hexagonal crystal of skyrmions forms spontaneously in a small region of the phase diagram positioned at a finite magnetic field close to the critical temperature $T_c$, historically known as the ``A''-phase \cite{doi:10.1126/science.1166767,Bauer2016,Yu2011,doi:10.1126/science.1214143}. 
The thermodynamic stability of skyrmions relies here on thermal fluctuations that are still pronounced close to $T_c$ \cite{doi:10.1126/science.1166767,Buhrandt2013,Laliena2017a}. If the size of SOC surpasses a critical threshold as in Cu$_2$SeO$_3$, magnetocrystalline and exchange anisotropies will stabilize an additional skyrmion phase that, importantly, extends down to zero temperature \cite{Chacon2018,Halder2018,Bannenberg2019}. Cubic magnets with octahedral point group $O$ like Co-Zn-Mn alloys with a $\beta$-Mn-type crystal lattice share basically the same magnetic phase diagram as the B20 materials \cite{Tokunaga2015}. As the spin-orbit coupling is relatively weak in all these materials, the size of skyrmions is rather large and typically ranges between 10-200 nm. An exception is the B20 compound MnGe where the heavier Ge atom leads to a much larger SOC and a modified phase diagram with magnetic textures characterized by 3 nm length scale. However, these textures contain hedgehog defects and are thus not skyrmionic \cite{Kanazawa2017a,Fujishiro2019}. 

Rhombohedral magnetic materials with polar point group $C_{3v}$  hosting N\'eel skyrmions are, with increasing strength of SOC, GaV$_4$S$_8$ \cite{Kezsmarki2015}, GaV$_4$Se$_8$ \cite{Fujima2017} and GaMo$_4$S$_8$ \cite{Butykai2022}. The latter two show skyrmion phases that are stable down to zero temperature for a magnetic field aligned with the polar $[ 111 ]$ axis. Replacing V$_4$ with heavier Mo$_4$ clusters increases the SOC, resulting in skyrmion sizes of approximately 10 nm. Finally, we mention that more exotic antiskyrmions are realized in certain Heusler compounds with $D_{2d}$ symmetry \cite{Nayak2017} and in the material 
Fe$_{1.9}$Ni$_{0.9}$Pd$_{0.2}$P possessing the point group $S_4$ \cite{Karube2021}. 

Synthesizing the above-mentioned bulk materials in thin film format reduces their symmetry, potentially stabilizing various phases. In particular, an additional uniaxial anisotropy $K$ is generated in thin films via the magnetic dipolar interaction, thus allowing phase tuning across a rich diagram which includes skyrmion phases that are stable down to zero temperature (Fig.~\ref{Fig:PhaseDiagrams}). Alternatively, magnetic monolayers or multilayers might be deposited on or combined with nonmagnetic metals with strong SOC, which generates an interfacial DMI \cite{Fert1980}. Such an optimized design can in turn stabilize skyrmions with a size of a few nanometers; see Refs.~\cite{Wiesendanger2016,Fert2017,Jiang2017} for reviews. 
 Notable examples include Ir/Fe/Co/Pt multilayers \cite{Raju2019,Raju2021} which offer a wide range of skyrmion sizes and stabilities via geometric tuning of the structure. However, compared to clean bulk materials such engineered magnetic layers are usually hampered by stronger disorder effects.

It should be noted that local DMIs can still emerge in globally centrosymmetric crystal structures, provided that the spontaneous magnetic moment which they induce is invariant under the action of all transformations of the magnetic symmetry class~\cite{Dzyaloshinsky1958}. This condition corresponds  to locally broken spatial inversion symmetry (e.g. within individual crystalline layers~\cite{Hayami2022a}). Consequently, DMIs with opposite chiralities may emerge on different spin sublattices, leading to exotic mixed-helicity skyrmionic textures~\cite{Hayami2022f,Lin2024,Cui2024}. One example of such behavior is the proposed formation of skyrmions with doping-dependent helicities in ferromagnetic van der Waals materials such as CrCl$_3$ and VCl$_3$~\cite{Tran2024}.

\subsubsection{Skyrmions in centrosymmetric crystals} 
\label{Sec:Tunable_Helicity}

In a centrosymmetric crystal, a global magnetic DMI with fixed chirality is not present as it is forbidden by symmetry. However, skyrmions may also be stabilized by frustrated symmetric exchange interactions \cite{Ivanov1990}, independently of the existence of any DMI. This scenario for skyrmion formation has been explored recently, with frustration either caused by the competition of nearest and next-nearest neighbor exchange interactions in combination with a triangular lattice \cite{Lin2016b,Martin2008,PhysRevLett.108.017206,Leonov2015,Hayami2022,Hayami2022b}, and/or the interplay between RKKY exchange~\cite{Ozawa2017,Takagi2018,Wang2020d,Bouaziz2022,Hayami2022,Hayami2022d,Hayami2023,Wang2023f} and other competing interactions (e.g. four-spin exchange~\cite{Heinze2011} or biquadratic spin correlations~\cite{Hayami2017}). Several recently discovered rare earth magnets with tetragonal~\cite{Khanh2020,Takagi2022}, triangular~\cite{Kurumaji2019}, and kagome~\cite{Hirschberger2019} lattices are skyrmion hosting materials, with RKKY being the dominant stabilization mechanism. Here, Fermi surface nesting results in a characteristic spin rotation length of twice the Fermi wavelength, $2\lambda_\mathrm{F}$. 

\begin{figure}[htbp]
\includegraphics[clip=true, width=0.99\columnwidth]{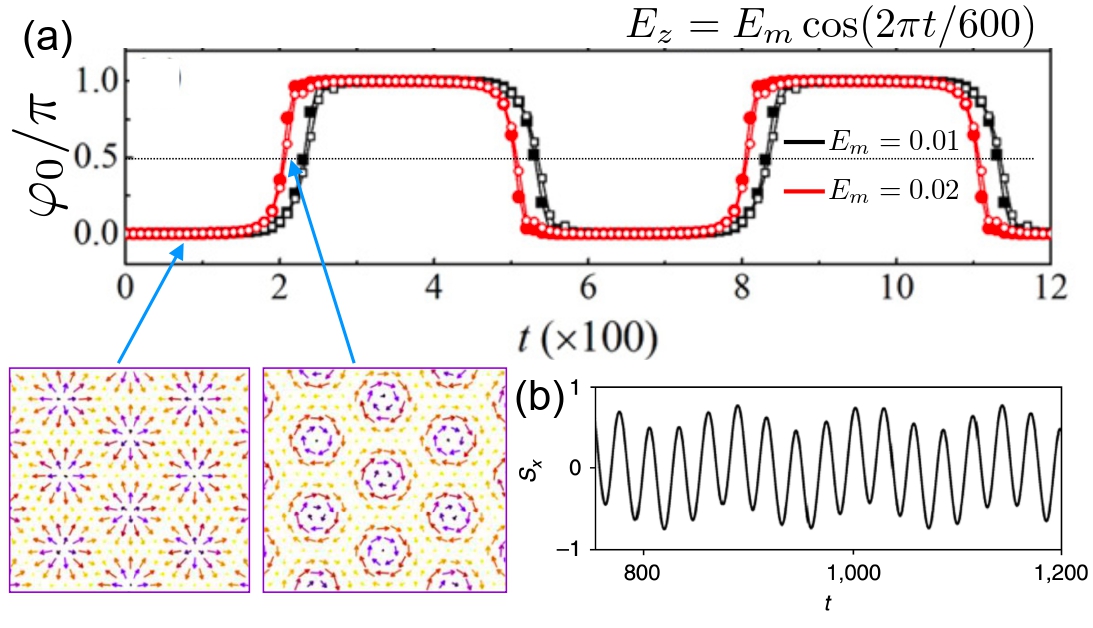}
\caption{\label{Fig4} Helicity tuning and dynamics. (a) The time dependence of skyrmion helicity $\varphi_0$ under the application of an AC electrical field along the z-axis, $E_z= E_m \cos(2\pi t/600)$ with $E_m=0.01$ (black points) and $E_m=0.02$ (red points). The lattice is modulated from N\'eel ($\vert E_z \vert> 0$) to Bloch ($\vert E_z \vert \approx 0$). b) Superposition of low-frequency helicity oscillations on the high-frequency rotation of the skyrmion center of magnetic mass, driven by an in-plane microwave excitation. Diagrams adapted from (a) \cite{Yao2020} and (b) \cite{Leonov2015}.}
\end{figure}

Importantly, in centrosymmetric systems (which lack a unique $\mathbf{D}$ vector and hence a single symmetry-controlled spin rotation axis), magnetic skyrmions stabilized by competing interactions are degenerated with respect to vorticity $m$ and helicity $\varphi_0$. Consequently, there exists a degenerate manifold of skyrmions and antiskyrmions. 
The continuous helicity corresponds to a zero mode, i.e., a low-energy degree of freedom that characterizes skyrmions stabilized by frustration. The helicity dynamics are thus expected to influence the motion of such skyrmions~\cite{Leonov2015,Lin2016b}. For example, in the presence of an in-plane microwave frequency magnetic field, a low-frequency counter-clockwise (CCW) rotation of the skyrmion helicity is superposed on the higher-frequency CCW rotation of its center of magnetic mass [Fig.~\ref{Fig4}(b)]. The coupling between helicity and center of mass modes is non-linear, evidenced by a non-linear relation between the drive amplitude and the helicity rotation frequency. Resonant dynamics of centrosymmetric skyrmions are hence more complex than in systems with a fixed spin chirality.    

These helicity oscillations may usefully impact the translational dynamics of a skyrmion, with simulations suggesting that the skyrmion Hall angle is suppressed for magnon-driven skyrmion motion~\cite{Jin2022}. This represents a significant advantage over the compromised linear skyrmion dynamics in chiral magnets: the fact that chiral skyrmions do not move parallel to applied currents is a major barrier to developing useful skyrmion devices, since spin torques cannot move skyrmions through narrow wires or channels without risking annihilation via interactions with the boundaries~\cite{Iwasaki2013a}.

If an out-of-plane electric field is applied, spatial inversion symmetry is broken and the skyrmion helicity switches to N\'eel 0 or $\pi$, depending on the field orientation~\cite{Yao2020}. This phenomenon originates from the intrinsic magnetoelectric effect in non-collinear insulating magnets, which is also responsible for the well-known inverse DMI~\cite{Katsura2005,Mostovoy2006}. Spin rotation with a specific chirality creates a spin current $\mathbf{j_s} \propto \mathbf{S}_i \times \mathbf{S}_j$, where $\mathbf{S}_{i,j}$ denote neighboring spins.  $\mathbf{j_s}$ spontaneously breaks spatial inversion symmetry: consequently, an electric dipole moment emerges with $\mathbf{P}\propto\mathbf{r_{ij}}\times\mathbf{j_s}$ (where $\mathbf{r_{ij}}$ is parallel to the wavevector of the spin helix).  The effect is maximal for N{\'e}el helicity; for Bloch skyrmions $\mathbf{r_{ij}} \parallel \mathbf{j_s}$ and hence no electrical polarization occurs. This principle is related to long-standing ideas of modifying magnetic order via an electrically-tuned Rashba SOC~\cite{Barnes2014}.

\section{
Quantum States in the Background of Classical Skyrmions}

Classical magnetization textures containing skyrmion configurations provide fertile ground for a rich variety of quantum excitations. These can emerge as intrinsic excitations in the form of magnons, i.e., quantized spin fluctuations, or via the interplay of electrons with skyrmions in metallic magnetic materials, substantially influencing transport characteristics. Quantum excitations may also be engineered using hybrid architectures. In particular, combining magnetic skyrmions with an $s$-wave superconductor provides a path to realize topological superconductivity~\cite{Nakosai2013}. Furthermore, cavity magnonics promises access to the strong coupling regime between photons and magnons of skyrmion textures. 

\subsection{Emerging quantum states in the presence of skyrmions} 
\label{sec:EmergentQuantumStates}

\subsubsection{Magnons of isolated classical skyrmions}

The small amplitude fluctuations of a classical magnetization texture are spin waves. In the lowest order in their amplitudes, they can be computed using linear spin-wave theory. Quantizing these fluctuations, one obtains the magnon excitations. As the skyrmion textures, similar to N\'eel antiferromagnet order, are in general not exact groundstates of the quantum Hamiltonian, these magnonic quantum fluctuations lead to renormalizations and may affect substantially the groundstate properties, see Fig.~\ref{Fig:Compendium}(e) and (f) for an illustration.
 
 The spin waves of a classical magnetization texture containing a single isolated skyrmion can in general be separated into bound states, whose spin wavefunctions are localized to the skyrmion, and extended scattering states. Depending on the resulting temporal evolution of the skyrmion configuration, the bound states can be classified as a breathing mode, CCW mode, quadrupolar mode, etc. \cite{Mochizuki2012,Iwasaki2014,Lin2014,Schuette2014}. 
 
 Interestingly, the scattering states are subject to a magnon Berry phase \cite{Dugaev2005}. When the spin wave scatters off a skyrmion, it collects an additional phase along its path, $\int^{\vec r} d\vec r' \vec a(\vec r')$, which is governed by the spin connection $a_i(\vec r) = \hat e_1 \partial_i \hat e_2$ where $\hat e_{1/2}$ are orthogonal unit vectors specifying the precessional plane orthogonal to the local magnetization $\vec m$, i.e., $\hat e_1 \times \hat e_2 = \vec m$. 
The associated Berry flux $(\nabla \times \vec a)_z = b_{\rm em}
 + 2\pi \sum_n s_n \delta(\vec r - \vec r_n)$ contains smooth and singular parts. The latter consists only of integer fluxes $s_n \in \mathbb{Z}$ located at certain positions $\vec r_n$, and is therefore physically unobservable and irrelevant. 
 The important smooth part $b_{\rm em} = 4\pi \rho_{\rm top} + \dots$ is governed by the topological charge of the texture $\rho_{\rm top}$ [Eq.~\eqref{TopChargeDensity}]. DMI may give rise to corrections to this smooth part, although these do not contribute to the total flux (at least for the cases known so far). As a result, the spin wave experiences an emergent magnetic field with total flux, $\int d\vec r\, b_{\rm em} = 4 \pi Q$, i.e. $4\pi$ per winding number $Q$. This emergent magnetic field causes skew scattering of magnons and a magnon Hall effect, for example, in heat transport \cite{Kong2013,vanHoogdalem2013,Mochizuki2014,Qin2022}. Real-space topology of classical textures therefore strongly influences the dynamics of quantum excitations. 

\subsubsection{Magnons of classical skyrmion lattices}

For a lattice of classical skyrmions, spin waves experience Bragg scattering, which leads to a magnon band structure $\omega_n(\vec k)$. Some of these modes at zero wavevectors $\omega_n(0)$ are similar to the bound spin waves of single skyrmions. Importantly, the breathing, CW, and CCW modes possess macroscopic magnetic dipole moments and can thus be probed in a resonance spectroscopy experiment with typical frequencies in the GHz range \cite{Mochizuki2012,Onose2012,Schwarze2015,Satywali2021}. The band structure $\omega_n(\vec k)$ is topologically non-trivial as well because many of the bands carry a finite Chern number. 

Each skyrmion with winding number $|Q| = 1$ provides an emergent magnetic flux of $4\pi$ per magnetic unit cell such that on average the spin waves experience a finite emergent magnetic field. The band structure can thus be interpreted as a collection of magnon Landau levels. The non-trivial topology of this magnon band structure in reciprocal space is inherited from the non-trivial topology of the skyrmion lattice in real space \cite{Roldan-Molina2016,Garst2017,Weber2022}. Due to this non-trivial topology, stable magnon edge states are expected with energies located within the gap separating two bands with distinct Chern numbers. By tuning the band structure using a magnetic field and other parameters, its topology can be altered, leading to the emergence or disappearance of the edge states \cite{Diaz2020}. The experimental detection of these edge states, however, has not yet been achieved.

\subsubsection{Electrons and classical skyrmion lattices}

In a metallic magnet, conduction electrons will be also influenced by the classical magnetic texture in a non-trivial manner \cite{Volovik_1987,Bruno2004,Binz2008}. Assuming a local exchange interaction between the magnetization and the magnetic moment of the conduction electrons, the electrons tend to align their magnetic moment with the local magnetic texture. This amounts to a geometric constraint similar to the case of spin waves and gives rise to an emergent magnetic field governed by the topological charge density $\rho_{\rm top}$ of Eq.~\eqref{TopChargeDensity}. For electrons, this field corresponds on average to a flux quantum $\phi_0 = 2\pi \hbar/e$ per magnetic skyrmion with $|Q| = 1$. This corresponds to rather large emergent magnetic fields $B_{\rm em} = \phi_0/A_{\rm sk}$ with an order of magnitude between 40 and 4000 Tesla for a magnetic unit cell $A_{\rm sk}$ ranging between 100 and 1 nm$^2$. Note, however, that the density $\rho_{\rm top}$ varies strongly in space and the spatial variations of the emergent field are commonly much larger than its averaged value $B_{\rm em}$.

The emergent field plays an analogous role to the external magnetic field in the classical Hall effect, deflecting electrons with an emergent Lorentz force and hence generating a transverse voltage. The resulting semiclassical topological Hall effect has been experimentally measured in a series of materials, e.g.~\cite{Neubauer2009,Kurumaji2019,Raju2021}.  A \emph{quantized} topological Hall conductance has not yet been observed: this would require an electron mean free path greater than the skyrmion diameter, and sufficiently low measurement temperatures $k_B T \ll \Delta$ where $\Delta$ is the gap in the electron band structure induced by the texture \cite{Hamamoto2015}. Due to the spatial variation of the emergent field, the Landau levels are dispersive such that $\Delta$ can be much smaller than the Stoner gap. The observation of a quantized topological Hall effect thus necessitates clean, low carrier density materials whose skyrmion lattice phase remains stable down to the lowest temperatures.

\subsection{Superconductor-skyrmion hybrid structures}

\begin{figure*}[htbp]
\includegraphics[clip=true, width=1.99\columnwidth]{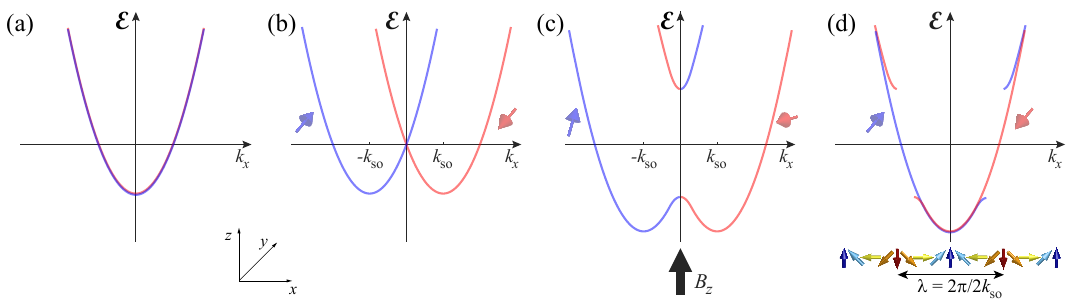}
\caption{\label{Fig5} Creation of a ``spinless'' Fermi surface for topological superconductivity. (a) The departure point: a typical spin-degenerate band dispersion $\varepsilon(k_x)$. (b) Rashba SOC (oriented along the spin $y$ axis) splits the dispersion along $k_x$ for opposite spin orientations. (c) Applying a uniform external Zeeman field along the $z$ axis breaks time-reversal symmetry, opening a gap at $k_x=0$. (d) The combination of orthogonal Rashba and Zeeman fields shown in (c) is gauge-equivalent to the spiral magnetic field generated by a real-space spin helix. The strength of this ``synthetic'' SOC scales inversely with the period of the spiral, while the orientation of the effective SOC vector is controlled by the spin helicity. Diagrams are adapted from refs.~\cite{Alicea2012,Braunecker2010}.}
\end{figure*}

\subsubsection{From trivial to topological superconductivity}\label{Sec:TScIntro}
Developing ``hybrid'' materials that interface a non-collinear magnet housing skyrmions with a conventional $s$-wave superconductor may provide an attractive route towards achieving topological superconductivity. A topological superconductor exhibits a gapless zero-energy bound state wherever its order parameter is suppressed to zero, i.e. at its phase boundaries and inside its vortex cores~\cite{Volovik2000,Read2000,Ivanov2001}. Due to particle-hole symmetry, these boundary quasiparticles are their own antiparticles, resembling Ettore Majorana's 1937 prediction of electrically-neutral self-conjugate particles described by a real-valued wave equation~\cite{Majorana1937}.  Majorana Zero Modes (MZM) in topological superconductors are anyonic quasiparticles obeying non-Abelian statistics under exchange and are central to proposals for the realization of topological qubits for quantum computing \cite{Nayak2008}.

MZM are predicted to emerge at each end of a one-dimensional superconducting wire with chiral $p$-wave pairing~\cite{Kitaev2001}. Creating a chiral $p$-wave order parameter necessitates inducing pairing between spinless fermions with opposite momenta. The first proposal for pairing spinless fermions used the spin-locked edge states of a topological insulator, proximitized by a $s$-wave superconductor~\cite{Fu2008}. Spinless pairing also occurs in proximitized 1D semiconducting nanowires~\cite{Lutchyn2010} and 2D electron gases~\cite{Sau2010}, in the presence of Rashba SOC and an orthogonal Zeeman field $B_Z$. This combination of interactions is replicated by helical magnetism (Fig.~\ref{Fig5}). 

The process of creating a spinless Fermi surface first uses SOC to break spin-degeneracy between opposite momenta [Fig.~\ref{Fig5}(b)], then an orthogonal Zeeman field to open a gap at the zero-momentum band crossing via spin-flip scattering [Fig.~\ref{Fig5}(c)]. Provided that the chemical potential $\mu$, pairing energy $\Delta_0$ and Zeeman splitting $\Delta_Z \equiv g^*\mu_B B_Z/2$ fulfill the topological criterion $\Delta_Z > \sqrt{\Delta_0^2+\mu^2}$~\cite{Alicea2012}, a chiral $p$-wave superconducting gap can be opened in the spinless Fermi surface via the proximity effect. The same mechanism and criterion holds for 2D Rashba superconductors in out-of-plane magnetic fields $B_z$, since the Rashba Hamiltonian $\mathcal{H}_R=(\mathbf{\sigma}\times\mathbf{k})\cdot\mathbf{\hat{z}}$ locks the spins in-plane, perpendicular to the electron momenta $\hbar\mathbf{k}$.  The topological nature of the superconducting order parameter is easier to perceive in 2D, since its phase winds continuously by $2\pi$ in the $(k_x,k_y)$ plane. Attempts to induce $p_x \pm ip_y$ pairing in topological insulators have been stymied by defect-induced charge leakage into bulk states.  Satisfying the topological criterion is also challenging in proximitized semiconductors, due to the large Zeeman fields required which risk quenching the parent superconductor.  

A 1D spin helix is gauge-equivalent to the recipe combining Rashba and Zeeman effects for a spinless Fermi surface~\cite{Braunecker2010,Braunecker2013,Klinovaja2013,Vazifeh2013,Steffensen2022}.  This is illustrated in Fig.~\ref{Fig5}(d), where the spin-dependent gauge transformation $\psi_\sigma(x) \rightarrow e^{i\sigma k_{so}x}\psi_\sigma(x)$ ($\psi_\sigma$ is the electron operator) removes the band shifts along the momentum axis and replaces the uniform Zeeman field $B_z$ with a spiral $\mathbf{B}(x)=B[\mathrm{cos}(2k_{so}x)\mathbf{\hat{z}}-\mathrm{sin}(2k_{so}x)\mathbf{\hat{x}}]$.  The spiral period $\pi/k_{so}=h/2m\alpha$, hence shorter helical wavelengths imply stronger effective SOC.  The new periodicity imposed by this spiral causes scattering with momentum transfer $\pm 2k_{so}$, opening a gap in the band structure at $\pm k_{so}$. However, the helicity of the spiral breaks spin symmetry, and therefore, only one of the two spin-dependent Fermi surfaces is gapped. One may therefore build a 2D $p_x \pm ip_y$ superconductor using a skyrmion lattice on the surface of a $s$-wave superconductor, exhibiting delocalized chiral Majorana edge states and MZM localized in vortex cores~\cite{Nakosai2013}. This intrinsically 2D architecture would greatly facilitate the "braiding" process for gate implementation, avoiding the demanding requirement to synthesize regular nanowire arrays and also promising MZM stabilization at or close to zero external magnetic field. Later analysis of similar heterostructures indicated that MZM formation would not require fine-tuning of the chemical potential $\mu$~\cite{Chen2015c}: a significant advantage over semiconducting nanowires for designing topological qubits. 

\subsubsection{Skyrmion-vortex pairing}

\begin{figure}[htbp]
\includegraphics[clip=true, width=0.99\columnwidth]{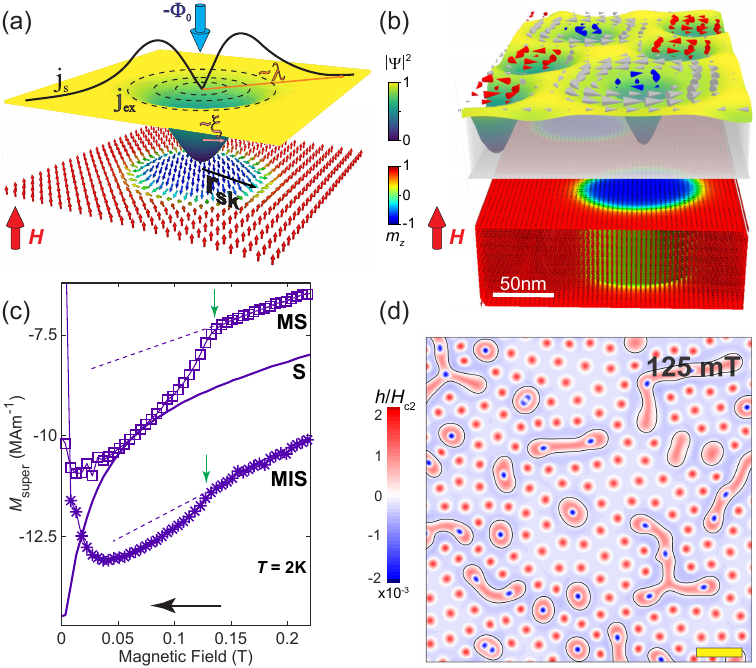}
\caption{\label{Fig6} Skyrmion-vortex pairing. (a) Schematic of antivortex nucleation by a N\'eel skyrmion. Strong coupling is ensured by engineering the material to satisfy the inequality $\xi<r_\mathrm{sk}<\Lambda$, where $\xi$ is the coherence length and $\Lambda$ the Pearl penetration length. (b) Ginzburg-Landau simulation of the superfluid and supercurrent densities above a micromagnetic reproduction of an experimentally-measured skyrmion. Red/blue arrows indicate vortex/antivortex supercurrents, respectively.  Vortex-antivortex annihilation is prevented by Meissner screening currents (gray arrows) whose density peaks at $r_\mathrm{sk}$. (c) Experimental detection of skyrmion-antivortex nucleation in chiral magnet/ultra-thin superconductor multilayers: the drop in the magnetization of the superconductor at the skyrmion nucleation field corresponds to the formation of antivortices. (d) Large-scale ($2\times2\,\mu$m) Ginzburg-Landau simulations of the out-of-plane flux density induced by the supercurrent profile in the same superconductor/chiral magnet system (yellow scale bar measures 250\,nm).  Skyrmion-antivortex pairs are visible as small blue dots encapsulated by red regions (created by the screening currents). Note the extremely low flux density associated with this array of Pearl vortices/antivortices. Data are reproduced from ref.~\cite{Petrovic2021a}.}
\end{figure}

The possibility of individual skyrmions creating localized MZM in a proximate superconductor was first addressed by~\cite{Yang2016b}, but the high-vorticity ($m =2$) skyrmions deemed necessary are not thermodynamically stable in a chiral magnet with DMI. An elongated skyrmion can generate an MZM at each end, with the MZM localization improved by increasing the SOC strength~\cite{Gungordu2018}. This idea merits future experimental attention, although the mobility of such elongated textures is limited and braiding would require a complex flux-linked device geometry similar to Rashba nanowires~\cite{Sau2011,VanHeck2012}. However, a skyrmion-vortex pair with $m+n=$\, even,  where $n$ is the number of flux quanta $\Phi_0$ in the vortex,  can spontaneously induce a localized MZM in the vortex core and a delocalized Majorana edge state at the system boundary~\cite{Rex2019}.

Skyrmions and vortices can interact via exchange (Rashba-Edelstein) coupling. This requires electrical contact between the superconductor and the magnet, in contrast to stray field (electromagnetic) coupling which is unaffected by an intermediate insulating layer. Rashba-Edelstein coupling relies on the principle that spin polarization $\mathbf{S}$ in a polar superconductor induces a supercurrent $\mathbf{J_S}$ following the relation $\mathbf{S} \propto \mathbf{\hat{z}} \times \mathbf{J_S}$, where $\mathbf{\hat{z}}$ is the polar axis (i.e. the normal to the plane at a 2D Rashba interface)~\cite{Edelstein1995}. 

Consequently, the exchange field $h_{ex}$ from the in-plane spin components of a N\'eel skyrmion will create a vortex-like supercurrent in a proximate superconductor~\cite{PhysRevLett.117.017001}. The supercurrent circulation direction is controlled by the skyrmion helicity $\varphi_0$ and polarity $p$, as well as the sign of the exchange field and SOC at the superconductor/magnet interface.  If the induced supercurrent density is sufficiently high, a vortex is generated to minimize the overall free energy of the hybrid~~\cite{Baumard2019}. The skyrmion-induced and vortex-induced supercurrents circulate in opposite directions: consequently, their spatial superposition minimizes the total supercurrent density in the superconductor.

Stray-field coupling in such hybrid systems can be thought of as the interaction between the dipole-like fields of the skyrmion and the vortex, whose magnitude scales with the magnetic flux associated with a single skyrmion~\cite{Dahir2019,Menezes2019}.  Above a critical value, the skyrmion induces a vortex in the superconductor (accompanied by an antivortex if the external field is sufficiently small). For N\'eel skyrmions in the weak-coupling limit (where the skyrmion flux $\ll\phi_0 \equiv h/2e$), switching $\varphi_0$ can lead to a sign-change in the Meissner screening current profile in the superconductor~\cite{Dahir2020}. 

These interactions can be simultaneously optimized by maximizing the skyrmion radius $r_\mathrm{sk}$ while ensuring that its size falls between the coherence and Pearl (penetration) lengths in the superconductor: $\xi<r_\mathrm{sk}<\Lambda$ [Fig.~\ref{Fig6}(a)], thus ensuring vortex and skyrmion supercurrent overlap. Following this principle, skyrmion-vortex pairs were successfully nucleated in a [Ir$_1$Fe$_{0.5}$Co$_{0.5}$Pt$_1$]$^{10}$/Nb$_{25}$ multilayer, specifically designed to exhibit the large DMI (2.1\,mJ/m$^2$) necessary to stabilize 50\,nm skyrmions at cryogenic temperatures~\cite{Petrovic2021a}. Although the skyrmion-vortex interactions were dominated by stray-field coupling, a weaker Rashba-Edelstein interaction component emerged at low temperatures ($T/T_c<0.5$), due to the reduced coherence length.

\subsection{Skyrmions and photons: Magnetoelectric cavity magnonics}

Quantum architectures based on magnetic systems coupled to quantum modules have surfaced as multitasking platforms with complementary functionalities \cite{Lachance-Quirion_2019}.  The fundamental ingredient for developing complex hybrid systems is a coherent interaction between magnetic modes and microwave/optical photons, phonons, or qubits. In a microwave cavity, magnetic moments $\mathbf{m}_{\mathbf{r}}$ interact with the magnetic field $\mathbf{B}(\mathbf{r})$ of the cavity modes via the usual magnetic dipole interaction $H_I = - \sum_{\mathbf{r}} \mathbf{m}_{\mathbf{r}} \cdot \mathbf{B}(\mathbf{r})$. In the second quantized form, it reads $H_I = \sum_n g_n (\hat{a}^{\dagger} \hat{c}_n+\hat{a} \hat{c}_n^\dagger)$,  where $\hat{a}^{\dagger}~(\hat{a})$ are the creation (annihilation) operators of a microwave photon,  $\hat{c}_n^{\dagger}~(\hat{c}_n)$ are the creation (annihilation) operators of a magnon mode $n$, and $g_n$ describe the coupling strength of each magnon mode to the microwave cavity mode. 

A hybrid system will enter the strong coupling regime when $g_n \gg \kappa_a, \kappa_n$, where $\kappa_a~(\kappa_n)$ is the cavity (magnon) mode linewidth.  In this regime cavity modes and resonant magnetic modes are hybridized, and coherent quantum information transfer is possible.  The linewidths of magnetostatic modes are proportional to the Gilbert damping $\alpha$ and lie in the MHz range, while $\kappa_a$ can reach the Hz range. Calculations indicate that the interaction between single photons and all resonant modes of a single topologically-trivial magnetic vortex in a soft ferromagnetic nanodisc can reach the strong coupling regime using materials with $\alpha \approx 10^{-3}$~\cite{Martinez-Perez_2019}, while the coupling to the skyrmion breathing mode lies one order of magnitude lower.  

In multiferroic materials, the electric polarization couples to the cavity's electric field as $H_{\mbox{\scriptsize EM}}=- \sum_{\mathbf{r}} \mathbf{p}_{\mathbf{r}} \cdot \mathbf{E}(\mathbf{r})$ and gives rise to a magnon-photon coupling with an electric $g_{\mbox{\scriptsize el}}$ and a magnetic component $g_{\mbox{\scriptsize mag}}$ \cite{PRXQuantum.3.040321}.  Among the magnon modes of a skyrmion lattice,  the CCW, CW, and breathing modes are magnetically active. In contrast, the elliptical modes are electrically active due to their magnetic quadrupole moment, for example, in the magnetoelectric insulator Cu$_2$OSeO$_3$ \cite{doi:10.1126/science.1214143,Kanazawa2017a}.

Estimates for the coupling strength in this material indicate that the strong coupling limit $g \approx 500$ MHz can be realized for the CCW and breathing modes, while a hybridization gap between the breathing and CCW modes arises as the result of the cavity-mediated magnon-magnon coupling. The photon-mediated magnon-magnon coupling between magnetoelectrically active skyrmion excitations in a cavity could potentially be used to generate entanglement between magnon qubits \cite{PRXQuantum.3.040321}. 
Notably, an experimental study of Cu$_2$OSeO$_3$ coupled to a microwave cavity resonator using magnetic resonance spectroscopy revealed a high magnon–photon cooperativity $C=g^2/(\kappa_n \kappa_a)$ in the skyrmion lattice phase \cite{PhysRevB.104.L100415}.  $C$ is highly tunable by small magnetic fields close to magnetic phase transitions: it increases from $C\approx 8$ in the helical phase to $C \approx 50$ in the skyrmion lattice, while remaining close to unity in the field-polarized and conical phases. 

\section{Quantum Skyrmions}

\subsection{Semiclassical quantization of skyrmions}

In the classical limit, magnetism is described by a unit vector field $\mathbf{m}$ that characterizes the local orientation of the magnetization. A common starting point for the formulation of a semiclassical approximation is the spin path integral representation \cite{assa}. Its central quantity is the action that for a two-dimensional system reads
\begin{align} 
\mathcal{S} = \int dt L = \int dt d^2 r [s \boldsymbol{\mathcal{A}}(\mathbf{m})\cdot \partial_t \mathbf{m} - \mathcal{H}(\mathbf{m})]\,.
\label{EuclideanAction}
\end{align}
Here, $\mathcal{H}$ is the spin Hamiltonian with all spin operators replaced by components of the unit vector field $\mathbf{m}(\vec r,t)$. In the continuum limit, $\mathcal{H}$ will not only depend on $\mathbf{m}$ but also on spatial derivatives, $\partial_i \mathbf{m}$ etc. The first term in Eq.~\eqref{EuclideanAction} is the Berry phase where $s$ is the spin density with units of $\hbar$ per volume. The vector potential is denoted by $\boldsymbol{\mathcal{A}}(\mathbf{m})$ that is not uniquely defined but it is only required to fulfill the relation $\mathbf{m} (\partial_{\mathbf{m}} \times \mathcal{A}) = 1$. As a consequence, the Euler-Lagrange equations associated with the action \eqref{EuclideanAction} just reproduce the classical Landau-Lifshitz equations, $\partial_t \mathbf{m} = - \gamma \mathbf{m} \times \mathbf{B}_{\rm eff}$ where the effective magnetic field is given by $\mathbf{B}_{\rm eff} = - \frac{1}{M_s} \frac{\delta}{\delta \mathbf{m}} \int d^2 r \mathcal{H}$ with the saturation magnetization $M_s = \gamma s$ and the gyromagnetic ratio $\gamma$.
As we will see, the size of the spin, i.e., the magnitude of $s$ controls the semiclassical approximation, and the classical limit is obtained for $s \to \infty$.

\subsubsection{Collective coordinates and their canonical quantization}

To formulate an effective low-energy theory for magnetic skyrmions, their relevant degrees of freedom characterizing the skyrmion dynamics need to be identified. For this purpose, it is useful to consider the symmetries of the system and the associated conserved quantities. The spatial translation symmetry and the associated conserved linear momentum play a particularly important role. 
It turns out that the linear momentum of a skyrmion is directly related to its spatial position. An important collective coordinate specifying the skyrmion's position is the first moment of its topological density $\rho_{\rm top}$, i.e., 
\begin{align} \label{FirstMoment}
\vec R = \frac{1}{Q} \int d^2 r\, \vec r \rho_{\rm top} \,,
\end{align}
where $Q$ is the skyrmion topological charge as defined in Eq.~\ref{TopCharge}. It can be shown that for the spatially confined topological density of a skyrmion texture, this collective coordinate satisfies the Poisson bracket \cite{PAPANICOLAOU1991425}
\begin{align} \label{PoissonBracket}
\{ R_i, R_j \} = \frac{\epsilon_{ji}}{4\pi s Q} \,,
\end{align}
where $\epsilon_{ij}$ is the antisymmetric tensor.
Comparing this result with the canonical Poisson bracket for position and linear momentum, $\{R_i, P_j\} = \delta_{ij}$, one obtains for the latter $P_i = 4\pi s Q \epsilon_{ji} R_j$. It can be verified that this linear momentum indeed corresponds to the generator of space translations, i.e., $\{P_i, \mathbf{m}_j\} =  \partial_i \mathbf{m}_j$.  

The equation of motion $\dot R_i = \{R_i , V(\vec R)\}$, where $V(\vec R)$ is an additional potential, acquires the form of the Thiele equation \cite{PhysRevLett.30.230} $\dot{\vec P} = {\bf G} \times \dot{\vec R} = {\bf F}$ where the gyrocoupling vector ${\bf G} = - 4\pi s Q \hat z$ and the force $F_i = - \partial V/\partial R_i$. This insight strongly constrains the form of the effective Lagrangian \cite{Stone1996},
\begin{align} \label{EffLagrangian}
L_{\rm eff}(\vec R, \dot{\vec R}) = 2\pi s Q \epsilon_{ji} \dot R_i R_j  - V(\vec R),
\end{align}
whose Euler-Lagrange equation reproduces the equation of motion above. For a translationally invariant system with vanishing potential $V(\vec R) = 0$ it follows $\dot {\vec R} = 0$, that is the skyrmion does not move. 

The result for the Poisson bracket \eqref{PoissonBracket} might be used as a basis for the canonical quantization of the skyrmion promoting position and momentum to operators, $\hat {\vec R}$ and $\hat {\vec P}$, obeying the commutator $[\hat R_i , \hat P_j] = i \hbar \delta_{ij}$ and $[\hat{R}_i,\hat{R}_j] = i\hbar \epsilon_{ji} / 4 \pi s Q$ \cite{Ochoa2019}.  The classical limit, where the two operators become commutative, is recovered when $s\rightarrow \infty$. 

\subsubsection{Quantum corrections and the skyrmion mass}

The semiclassical quantization discussed above relies on the algebraic structure of the Poisson bracket \eqref{PoissonBracket}. For the notion of a quantum skyrmion it is thus crucial whether this structure and the associated form of the Lagrangian \eqref{EffLagrangian} is robust with respect to quantum fluctuations. 
It has been suggested that such fluctuations change the structure and, in particular, generate a skyrmion mass $M$, i.e., an additional term in the Lagrangian quadratic in velocity \cite{Makhfudz2012,Ochoa2019}. A skyrmion mass has been widely discussed in the literature, and it commonly arises for other definitions of the collective coordinate $\vec R$ that are distinct from Eq.~\eqref{FirstMoment} \cite{Kravchuk2018,Tchernyshyov2022}. 

It is important to notice that even the definition \eqref{FirstMoment} for the collective coordinate acquires corrections due to quantum fluctuations. 
Small-amplitude quantum fluctuations correspond here to magnon excitations on top of the classical skyrmion solution that can be systematically analyzed in terms of a $1/s$ expansion. Expanding the first moment of topological charge \eqref{FirstMoment} in $1/s$ we obtain 
\begin{align}
\vec R = \vec R^{(0)} + \vec R^{(1)} + \mathcal{O}(s^{-1}).
\end{align}
The term $\vec R^{(0)}$ in zeroth order represents the collective coordinate of the skyrmion on the classical level. The lowest order correction of order $1/\sqrt{s}$ is attributed to a cloud of magnons, whose position is represented by $\vec R^{(1)}$. As the conservation law for a translationally invariant system demands that $\dot{\vec R} = 0$, it follows that this magnon cloud is bound to the classical skyrmion coordinate in a manner such that $\dot{\vec R}^{(0)} = - \dot{\vec R}^{(1)}$ in lowest order in $1/s$. 

The collective coordinate Eq.~\eqref{FirstMoment} reduces to the classical limit $\vec R^{(0)}$ when it is evaluated with a rigid skyrmion profile $\vec m_0(\vec r - \vec R^{(0)}(t))$. An effective theory for $\vec R^{(0)}$ has been derived by systematically integrating out magnon fluctuations \cite{PhysRevX.7.041045}, 
\begin{align} \label{MassiveLagrangian}
L_{\rm eff}(\vec R^{(0)}, \dot{\vec R}^{(0)}) = 2\pi s Q \epsilon_{ji} R^{(0)}_i \dot R^{(0)}_j + \frac{M}{2} \dot R^{(0)}_i \dot R^{(0)}_i + \dots
\end{align}
leading to a finite skyrmion mass $M$. This Lagrangian describes a massive charged particle in a magnetic field, $B = 4\pi s |Q|/q$ with the charge $q$, where the vector potential is given in the symmetric gauge, i.e., $q A_j = 2\pi s Q \epsilon_{ji} R^{(0)}_i$. 
Due to a finite mass $M$, new solutions of the equation of motions for $\vec R^{(0)}$ arise that correspond to classical cyclotron orbits with frequency $\omega_c = |G|/M = 4\pi s |Q|/M$. These orbits reflect the revolving motion of the classical $\vec R^{(0)}$ and the magnon cloud around a constant collective coordinate $\vec R$.  Note that the emergent magnetic field associated with a finite topological winding $Q$ is a theme common to the dynamics of skyrmions, magnons and electrons, see section \ref{sec:EmergentQuantumStates}.

As it corresponds to a conserved quantity, the first moment of topological charge $\vec R$ is the proper hydrodynamic coordinate. Is the classical coordinate $\vec R^{(0)}$ then physically relevant at all and could the frequency $\omega_c$ be observed experimentally? This seems to hinge on the existence of an experimental observable that couples to $\vec R^{(0)}$ rather than $\vec R$, which would eventually allow to probe the massive dynamics of Eq.~\eqref{MassiveLagrangian}.


\subsubsection{Quantum skyrmion states: gas, liquid, and crystal}

What is the ground state of a magnet containing skyrmions? The spectrum of the theory \eqref{EffLagrangian} for the collective variable $\vec R$ in the absence of a potential $V(\vec R)$ consists of a single highly degenerate level. The theory \eqref{EffLagrangian} has the same form as \eqref{MassiveLagrangian} in the zero mass limit where the cyclotron frequency $\omega_c \to \infty$. Correspondingly, the single highly degenerate level shares similarities with the lowest Landau level (LLL) for which the skyrmion wavefunctions are strongly localized in a region with linear dimension given by the magnetic length $\ell = \sqrt{\hbar/|G|} = \sqrt{\hbar/(4\pi s |Q|)}$. This reflects the zero-point motion of the skyrmions due to the Heisenberg uncertainty associated with the commutator $[\hat{R}_i,\hat{R}_j] = i\hbar \epsilon_{ji} / 4 \pi s Q$. 
Disorder breaks the translational invariance and might give rise to a pinning potential $V(\vec R)$. As a result, the degeneracy of the LLL is lifted, and the spectrum becomes equidistant with an energy separation $\Delta E \propto U_0(\ell/ \ell_p)^2$, where $U_0$ and $\ell_p$ characterize the strength and the size of the pinning potential, respectively \cite{PhysRevB.88.060404}.

In a clean system, the underlying atomic lattice will also provide an effective potential $V(\vec R)$ for the skyrmion coordinate that will possess the periodicity of the atomic crystal. For the theory \eqref{EffLagrangian}, the effective Hamiltonian is just given by $H_{\rm eff} = V(\vec R)$, and, in this case, the spectrum can be derived from a set of Harper's equations \cite{PhysRevB.94.134415,Ochoa2019}. The LLL is here split into a set of dispersive bands $\mathcal{E}_{n,\mathbf{k}}$ with $n= 0, ..., \vert 2 S Q \vert$ the band index and $S$ the total spin of the unit cell, $S=s a^2/\hbar$, for a square lattice with lattice spacing $a$. The band splitting $\Delta E$ as well as the dispersion are controlled by the effective strength of the atomic crystal potential that strongly depends on the ratio between the linear size of the skyrmion, $r_{\rm sk}$, and the lattice spacing $a$, $\log \Delta E \propto - (r_{\rm sk}/a)^2 $. Consequently, the dispersion is only appreciable if both possess the same order of magnitude, i.e., if the skyrmion size is comparable to the underlying lattice spacing. Note that $\Delta E$ is much smaller than typical magnon energies in the microwave range. 

In this limit, skyrmions behave as quasiparticles with a nontrivial band structure. At large field, this skyrmion quasiparticle is an excitation of the field-polarized ground state and their density will be dilute giving rise to a skyrmion gas. As the magnetic field is lowered, the excitation energy of skyrmions decreases and they might eventually condense. This process depends, however, on the precise form of their dispersion that in turn depends on the parameter $2 S Q$ \cite{PhysRevB.94.134415}. 
On the one hand, for odd $2 S Q$, a quantum phase transition is expected at the first critical field between the field-polarized state and a skyrmion liquid that is in the universality class of Bose-Einstein condensation. On the other hand, for even $2 S Q$ there is no sharp transition but only a crossover into the skyrmion liquid phase. For even lower magnetic fields, skyrmions will crystallize into a skyrmion crystal at a second critical field in agreement with the classical analysis. The intermediate skyrmion liquid phase, which is well-defined for odd $2 S Q$, is thus a genuine quantum effect.

The band structure, arising from the solution of the skyrmion Hamiltonian $H_{\rm eff}$, also possesses a nontrivial topology characterized by integer-valued Chern numbers $C_n$. As a consequence, for energies within the band gap counter-propagating chiral modes might appear, localized at the boundaries of a confined geometry, whose number and chirality are dictated by $C_n$ \cite{Ochoa2019}.   

\begin{figure*}[t]
\centering
\includegraphics[scale=0.4]{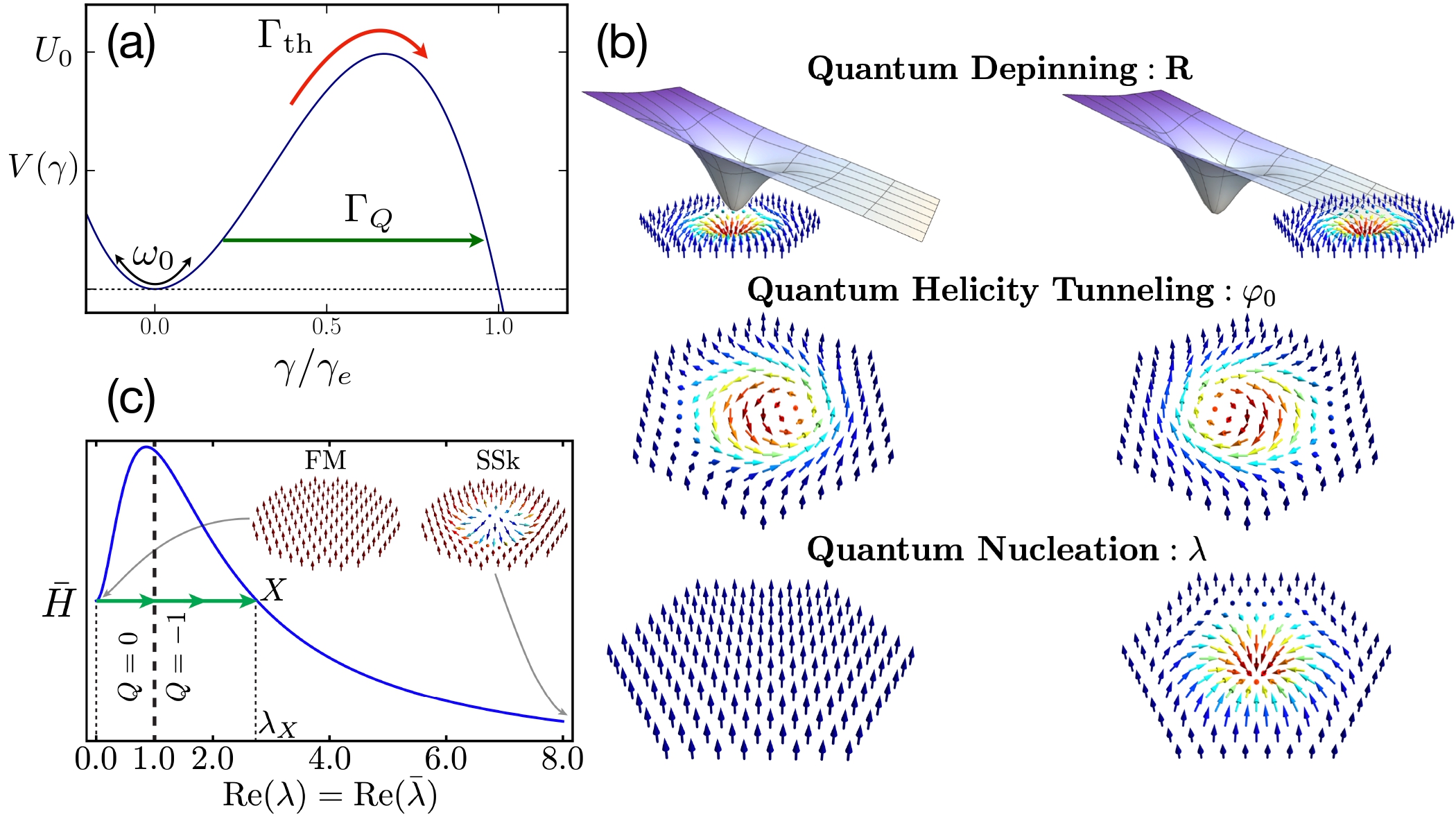}
\caption{Macroscopic quantum tunneling. a) A magnetic skyrmion is described in terms of a collective coordinate $\gamma$ prepared at the minimum of the potential $V(\gamma)$ of height $U_0$. Quantum tunneling through the potential barrier (green line) is possible as long as the inverse quantum rate $\Gamma_Q^{-1}$ can be observed on a laboratory time scale and the critical temperature above which thermal activation $\Gamma_{\mbox{\scriptsize th}}$ takes place is within experimental reach. b) Macroscopic quantum processes involve the quantum depinning of the skyrmion position $\mathbf{R}$ out of a defect, tunneling between two macroscopic skyrmion states with distinct helicities and quantum nucleation or annihilation processes. c) Energy landscape as a function of the collective coordinates $\lambda$ describing the quantum nucleation of a single skyrmion from the ferromagnetic state. Diagram adapted from \cite{doi:10.1142/9789811231711_0004}.}
\label{Fig:panel_tunneling}
\end{figure*}

\subsection{Macroscopic quantum tunneling}

This section aims to describe quantum behavior, emerging when many elementary magnetic moments forming a skyrmion coherently tunnel from a metastable configuration to a more stable state. Quantum tunneling within a macrospin approximation has been extensively studied and directly observed in the context of molecular magnets \cite{doi:10.1126/science.258.5081.414}, and is expected to be common in magnetic skyrmions. The problem is treated in terms of arbitrary collective coordinates $\gamma$ tunneling out of local minima that can describe different phenomena. We consider a skyrmion in the potential of Fig.~\ref{Fig:panel_tunneling}-a) prepared in a metastable state of low-lying energy levels near $\gamma=0$ in thermal equilibrium with a bath of temperature $T$.  The ground state energy is small, $\hbar \omega \ll U_0$,  with $\omega_0$ the frequency of small oscillations around the metastable minimum and $U_0$ the barrier height. Under this condition, quantum transitions are rare and the system behaves mostly classical. Here, our focus lies on the tunneling behavior of the collective coordinate $\gamma$ which resembles a single particle under a barrier. The problem of tunneling of the magnetization field $\mathbf{m}$ can also be studied under certain conditions \cite{PhysRevLett.60.661}. 

Thermally-assisted overbarrier transitions occur with an escape rate $\Gamma_{\mbox{\scriptsize th}}=(\omega_0/2\pi) e^{-U_0/k_B T}$: this classical behavior has been observed by resonant ultrasound spectroscopy of current-driven skyrmions in MnSi~\cite{Luo2020a}. At zero temperature, the Wentzel-Kramers-Brillouin (WKB) formula gives a quantum mechanical decay rate of $\Gamma_Q= 4 \omega_b \sqrt{15 \mathcal{S}_0 /2\pi} e^{-\mathcal{S}_0}$ \cite{weiss2008quantum}. Several generalizations of $\Gamma_{\mbox{\scriptsize th}}$ that incorporate the effect of damping can be found in \cite{weiss2008quantum}. Here $\mathcal{S}_0 = \mathcal{S} [\gamma_b]$ is the tunneling action calculated at the stationary instanton trajectory $\gamma_b(\tau)$ with a characteristic tunnel frequency $\omega_b$. The WKB exponent is $\mathcal{S}_0 \propto U_0/\omega_b \propto \gamma_e \sqrt{U_0/M}$, with $\gamma_e$ the tunneling distance.  The decay rate becomes determined solely by quantum effects below a characteristic temperature that marks the crossover from quantum tunneling to purely thermal activation, $T_c \approx U_0/k_B \mathcal{S}_0$.  To make quantum effects observable,  $\Gamma_Q^{-1}$  must not exceed a few hours and $T_c$ must be within experimental reach.  
 
 \subsubsection{Macroscopic quantum depinning}
 
The interaction of a skyrmion with a defect is sufficiently understood in the classical limit \cite{PhysRevB.91.054410,RevModPhys.94.035005}. The quantum depinning of a magnetic skyrmion out of an impurity potential created by an atomic defect in a mesoscopic magnetic insulator was studied in \onlinecite{PhysRevLett.124.097202} [see Fig.~\ref{Fig:panel_tunneling}-b)].  The Lagrangian of the collectives coordinates $\mathbf{R}$ is given by Eq.~\ref{EffLagrangian} with $V(R_x,R_y) = V_p(R_x,R_y) - F R_x$, where $V_p$ is the effective pinning potential of height $U_0$ and size $\lambda_D$, and $F$ is a linear force as the result of an applied out-of-plane magnetic field gradient. For small barrier heights, the Magnus force dynamics dominate over the inertial term, and the problem is reduced to a massless charged particle in a strong magnetic field.  The 2D instanton trajectory $\mathbf{R}_b$ in imaginary time is found by the requirement $V(R_b^x,i R_b^y)=0$,  in which $R_b^x$ ranges from zero up to the turning $R_d$ denoting the escape of the skyrmion into the classically forbidden region along the direction of the magnetic field gradient. Due to the two-dimensionality of the instanton path, $\mathcal{S}_0$ depends only on the width of the pinning potential $\lambda_D$ but is found to be independent of the potential height. Tunneling estimates for the magnetic insulator Cu$_2$OSeO$_3$ \cite{doi:10.1126/science.1214143} indicate that for sufficiently small skyrmions with a radius of a few lattice sites, coherent tunneling out of a pinning potential is expected to take place within a few seconds $\Gamma_Q^{-1} \approx$ s  in the millikelvin temperature regime $T_c \approx$ 100 mK.  Thus, magnetic skyrmions consisting of some thousands of spins can behave as quantum objects and display a behavior already observed in similar mesoscopic particles \cite{Brooke2001,PhysRevB.85.180401}.  
 
\subsubsection{Quantized skyrmion helicity}\label{Sec:MQT_Helicity}

As discussed in section \ref{Sec:Tunable_Helicity}, skyrmions in frustrated magnets possess a low-energy helicity degree of freedom $\varphi_0$ \cite{Leonov2015,Lin2016b}. In the presence of spin-rotation symmetry the helicity is a zero mode which can be quantized semiclassically. By perturbatively breaking the spin-rotation symmetry, an effective potential can be generated for the helicity, which favours certain discrete values for $\varphi_0$. The helicity dynamics in turn might lead to quantum tunneling between these discrete helicity values. 

By employing quantum field theory methods \cite{PhysRevB.106.104422}, the skyrmion dynamics is derived in terms of $\varphi_0(\tau)$ governed by the imaginary time Euclidean action,
\begin{align}
\mathcal{S}_E=\int_0^\beta d\tau~[\frac{\mathcal{M}}{2}\dot{\varphi}_0^2-iA\dot{\varphi}_0+V(\varphi_0)]\,.
\end{align}
Here $\mathcal{M}$ is the helicity mass, the potential 
is in general given by $V(\varphi_0)=V_0 \cos 2 \varphi_0 -V_1 \cos \varphi_0+V_2 \sin \varphi_0$,  and $A=h \mathcal{M}/s -s \Lambda$ with the magnetic field $h$ is a gauge field which does not affect the classical equations of motion.  $V_0$ is the result of dipole-dipole interaction \cite{Zhang2017} or in-plane uniaxial anisotropy induced by stress \cite{Shibata2015}. The application of an electric field produces $V_1$ and provides a direct external parameter to tune the minimum of the potential and thus the energetically favoured skyrmion helicity \cite{Yao2020}, while the last term $V_2$ corresponds to the application of an external magnetic field gradient or the presence of a DMI term. The quantity $s\Lambda= s \int d r^2 [1-m_z^0]$ is roughly the spin of the skyrmion, with $m_z^0$ the static skyrmion profile. The model resembles an electron moving on a conducting ring crossed by flux where $A$ is the vector potential of the magnetic flux penetrating the ring.  

Depending on the potential landscape $V(\varphi_0)$,  three distinct cases of quantum tunneling are possible: (i) macroscopic quantum tunneling (MQT), (ii) macroscopic quantum coherence (MQC), and (iii) macroscopic quantum oscillation (MQO).  For the former case, the escape tunneling rate is given by the standard WKB expression with $\Gamma_Q^{-1}\approx$ s and $T_c \approx 100$ mK, while the WKB exponent $\mathcal{S}_0$ grows exponentially with the skyrmion size $\lambda$ and the effective spin $S$.  When the skyrmion helicity experiences a double well ($V_2=0$) with two degenerate levels,  depicted in Fig.~\ref{Fig:panel_skyrmion_qubit} (a), tunneling between the two minima lifts the degeneracy and gives rise to an energy tunnel splitting $\Delta E =16 \omega_b \sqrt{U_0 \varphi_s /\pi \mathcal{M} \omega_b} e^{-\mathcal{S}_0}$, in the MHz regime, with $\varphi_s$ the distance between the two well minima.  

Finally,  tunneling between the well minima of a periodic potential ($V_1=0=V_2$) can occur either CW or CCW with an action of the form $\mathcal{S}_E = \mathcal{S}_0 \pm i A \pi$,  with the gauge term $A$ giving rise to a phase with an opposite sign, depending on the direction of tunneling. The energy splitting is given by $\Delta E' =\vert \cos (\pi A) \vert \Delta E$ and is quenched whenever $A=h \mathcal{M}/s -s \Lambda=n+1/2$, with $n$ an integer.  This is analogous to the spin-parity effect in magnetic systems with half-integer spin, related to the quantum tunneling suppression as a result of destructive quantum interference between tunneling paths \cite{PhysRevLett.69.3232}.  An additional tunneling rate quenching with the external field is also present, related to an Aharonov-Bohm type MQO in spin systems \cite{PhysRevB.61.8856}. 

\subsubsection{Quantum nucleation and collapse of skyrmions}

Individual magnetic skyrmions can be nucleated in a controllable way using among others, spin-polarized electrons generated by a scanning tunneling microscope \cite{doi:10.1126/science.1240573},  geometrical confinement under magnetic field protocols \cite{doi:10.1073/pnas.1600197113} and ultrashort single optical laser pulses \cite{PhysRevLett.110.177205}.  Theoretically, skyrmion nucleation processes from the ferromagnetic state are only allowed when spins are defined on a lattice,  $\mathbf{m}(\mathbf{r}) \rightarrow \mathbf{m}_{\mathbf{r}}$, where the continuum field theory and the notion of topological protection break down.  The zero temperature quantum nucleation of a single skyrmion from the ferromagnetic phase was studied in Ref.~\onlinecite{doi:10.1142/9789811231711_0004} and is driven by a local magnetic field over a circular spot. 

The magnetic phase $\mathbf{m}_\mathbf{r}$ is parametrized by a collective coordinate $\lambda(\tau)$ such that for $\lambda\rightarrow 0$, $\mathbf{m}$ describes the ferromagnetic state, while for $\lambda \rightarrow\infty$, $\mathbf{m}$ is a single skyrmion.  The energy $H(\lambda)$ has a local minimum at $\lambda=0$ and a global one at $\lambda =\infty$ [see Fig.~\ref{Fig:panel_tunneling}-c)]. The system quantum mechanically tunnels through the barrier from the ferromagnetic state to an intermediate state $X$ with $Q=-1$, and the single skyrmion state of just a few lattice sites is then reached via classical evolution.  As expected, the tunneling rate $\Gamma_Q$ strongly depends on the number of spins flipped during the tunneling process and decreases with increasing local magnetic field radius and strength. 

The reverse process involves the quantum collapse of a skyrmion due to the discreteness of a crystal lattice\cite{PhysRevB.98.024423,PhysRevB.86.024429}. In this physical scenario, zero-temperature skyrmion collapse occurs via underbarrier quantum contraction processes that conserve the topological charge but change the skyrmion size and magnetic moment. For nanometer-size skyrmions, these processes occur on a nanosecond time scale, as long as the external magnetic field $H$ is close to the critical field $H_c$ below which stable skyrmions can exist, i.e.  $\epsilon \ll 1-H/H_c$. The skyrmion then continues to contract until it reaches an atomic size and disappears by changing its charge from $Q=-1$ to $Q=0$.  The calculation of quantum tunneling parameters for realistic materials reveals that the lifetime of small skyrmions $\Gamma_Q^{-1}$ is on a laboratory time scale of seconds or minutes at the critical temperature $T_c \approx 1-4$ K \cite{Vlasov_2020}.

\subsubsection{Experimental implications}

Direct observations of macroscopic quantum tunneling suggest that quantum effects are common among mesoscopic spin systems. Typically, quantum tunneling has been investigated in high-spin single-molecule magnets and is observed by magnetization hysteresis with a staircase structure \cite{PhysRevLett.76.3830}. Both hysteresis loops \cite{Wernsdorfer2002} and magnetic relaxation time \cite{PhysRevLett.78.4645} become temperature-independent below a crossover temperature of $300-400$ mK. Quantum coherent tunneling has been observed through frequency-dependent magnetic susceptibility and magnetic noise measurement with a dc SQUID susceptometer below 200 mK\cite{doi:10.1126/science.258.5081.414}. The tunneling rate quenching originating from the destructive interference of tunneling paths was investigated using an array of micro-SQUIDs in single crystals of Fe8 \cite{doi:10.1126/science.284.5411.133}. Experimental evidence of quantum tunneling of a domain wall \cite{Brooke2001} and a magnetic vortex \cite{PhysRevB.85.180401} through pinning barriers suggests these processes occur at experimentally accessible temperatures. 

 \subsection{Skyrmion qubit}

 \begin{figure}[t]
	\centering
	\includegraphics[width=1\linewidth]{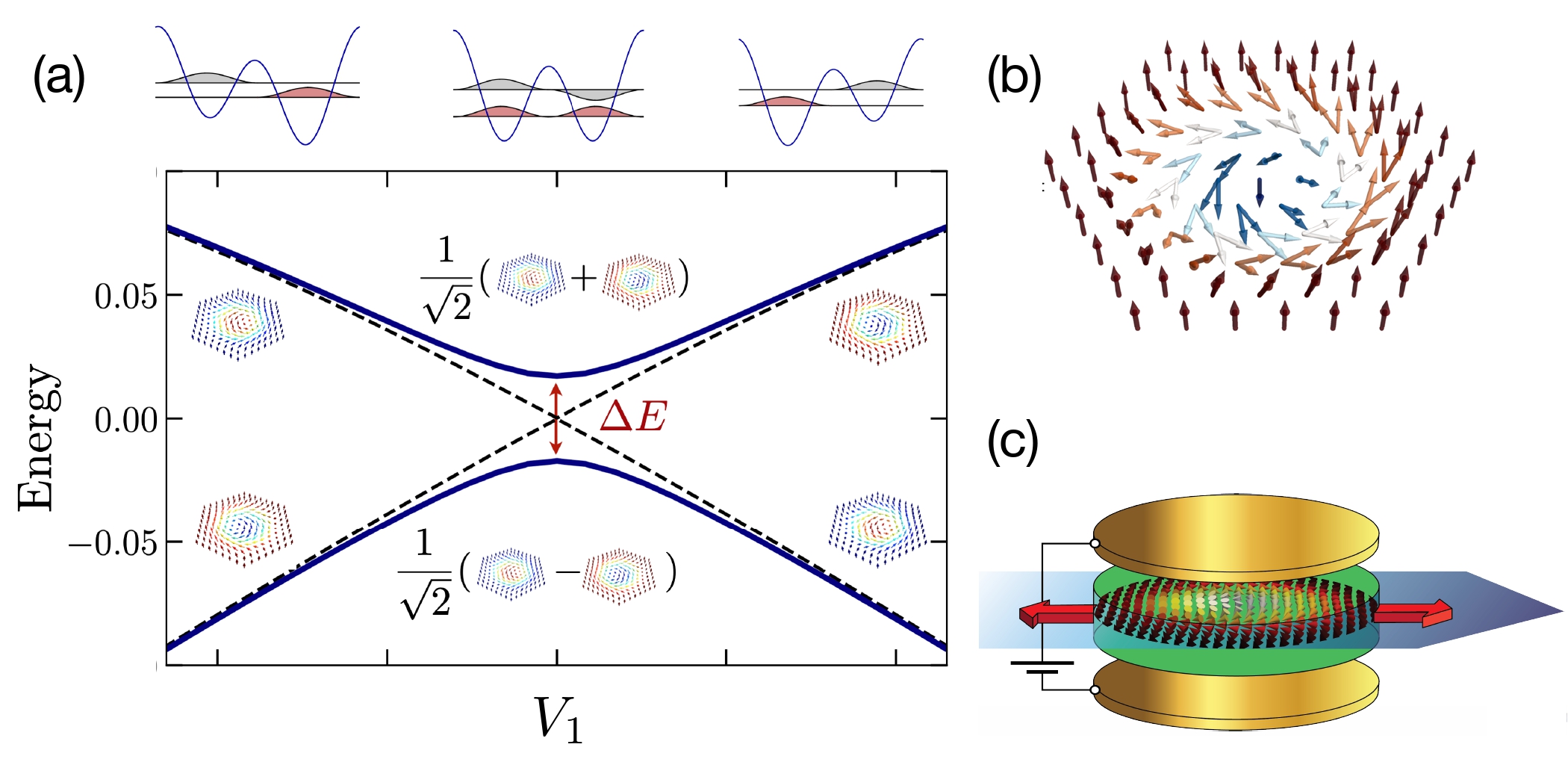}
	\caption{Skyrmion qubits. a) Schematic anticrossing. A skyrmion experiences a double-well potential separating two macroscopic states with distinct helicities forming a qubit. Close to the degenerate point, macroscopic tunneling lifts the degeneracy and eigenstates correspond to coherent superpositions of the two macroscopic states (b), separated by an energy-splitting $\Delta E$. c) The energy barrier can be varied in situ in the experiment using an electric field (yellow plates), uniaxial tensile stress (red arrows), and an in-plane magnetic field gradient (blue arrow). Schematics adapted from (b) \cite{PhysRevLett.127.067201} and (c) \cite{PhysRevB.106.104422}.}
	\label{Fig:panel_skyrmion_qubit}
\end{figure}

The potential for quantum tunneling between two macroscopic states with distinct helicities opens up new opportunities to leverage the quantum dynamics of magnetic skyrmions for the development of novel macroscopic qubits for quantum computing \cite{PhysRevLett.127.067201,10.1063/5.0177864}.  We focus here on the conceptual framework and key components for harnessing the quantum properties of skyrmion helicity, depicted in Fig.~\ref{Fig:panel_skyrmion_qubit}. A qubit using a skyrmion is created based on the quantization of energy levels associated with the helicity degree of freedom $\varphi_0$ and its conjugate momentum $\hat{P}$ associated with global spin rotations, $\hat{P} = \int dr^2 (1-m_z)-\Lambda$ satisfying $[\hat{\varphi}_0,\hat{P}] =i/S$. The corresponding quantum Hamiltonian is $\hat{H} =\hat{P}^2/2\mathcal{M} + h \hat{P} +V(\varphi_0)$, where the details of the various parameters are presented in Sec.~\ref{Sec:MQT_Helicity}.  Skyrmion qubits utilize skyrmion helicity’s energy-level quantization to encode quantum information as shown in Fig.~\ref{Fig:panel_skyrmion_qubit}. They are mapped on a simple physical basis of two macroscopic states based on either eigenstates of $\hat{P}$, quantized magnetic excitations of the perpendicular to the 2D plane magnetization component, or $\hat{\varphi}_0$ states with a well-defined helicity.  

The concept of skyrmion qubits offers a significant advantage by enabling the manipulation of energy-level spectra through external parameters, providing a versatile operating regime with high anharmonicity and tunable properties.  Electric fields \cite{Hsu2017} and magnetic field gradients\cite{Casiraghi2019} have emerged as a powerful tool for a current-free control of skyrmions and provide a direct way for tuning skyrmion helicity.  In addition, helicity can be manipulated with currents \cite{Zhang2017}, rendering the application of spin torques a new tool for the control of quantum information \cite{Sutton2015}. Skyrmion qubits are susceptible to environmental noise which inevitably results in quantum decoherence, a significant challenge in skyrmion-qubit devices. The important timescales for the qubit decoherence are the energy relaxation time $T_1$ and dephasing time $T_2$, which can be estimated starting from the magnetization dynamics encoded in the LLG\cite{1353448}. For sufficiently small skyrmions with radius $\lambda \approx 5-10$nm,  a modest choice of effective spin $S \approx 1-10$, an ultra-low Gilbert damping $\alpha=10^{-5}$ and low operational temperature $T = 100 $ mK, both $T_1$ and $T_2$ are estimated to be in the microsecond regime \cite{PhysRevLett.127.067201}.

Quantum algorithms can be implemented by a small set of single-qubit and two-qubit unitary operations.  Microwave magnetic field gradients can be used to drive single-qubit rotations around the $x$ and $y$ axis, $U_{x,y}$ \cite{PhysRevLett.127.067201}. Rotations around $z$, known as virtual zero-duration $U_z$ gates \cite{PhysRevA.96.022330}, are generated by adjusting the phase of the drive. Time-dependent electric field protocols generate $U_z$, while a time-dependent spin current generates $U_x$ \cite{PhysRevLett.130.106701}.  Thus, $U_{x,y,z}$ gates are simple to implement and natively available in a skyrmion quantum processor. 

The bilayer interaction translates into a two-qubit Hamiltonian with both transverse and longitudinal couplings. The two-qubit Ising interaction yields the CNOT and CZ gates, commonly used points of reference in the creation of quantum circuits, employing one two-bit operation under a time-dependent skyrmion bilayer interaction protocol \cite{PhysRevLett.130.106701}. Magnetic skyrmions hold promise as a new qubit type based on skyrmion helicity, extending their suitability from the classical to the quantum regime.  Although the potential for competitive technology exists, developing a skyrmion quantum processor faces significant challenges, including the availability of low-damping quantum skyrmion-hosting materials, the creation of clean interfaces, and the scalability of device architecture \cite{10.1063/5.0177864}. In Sec.~\ref{SubSec:Materials} and Sec.~\ref{SubSec:Devices} we discuss some of the challenges in developing quantum processors based on skyrmions.

\begin{figure*}[t]
	\centering
	\includegraphics[width=1\textwidth]{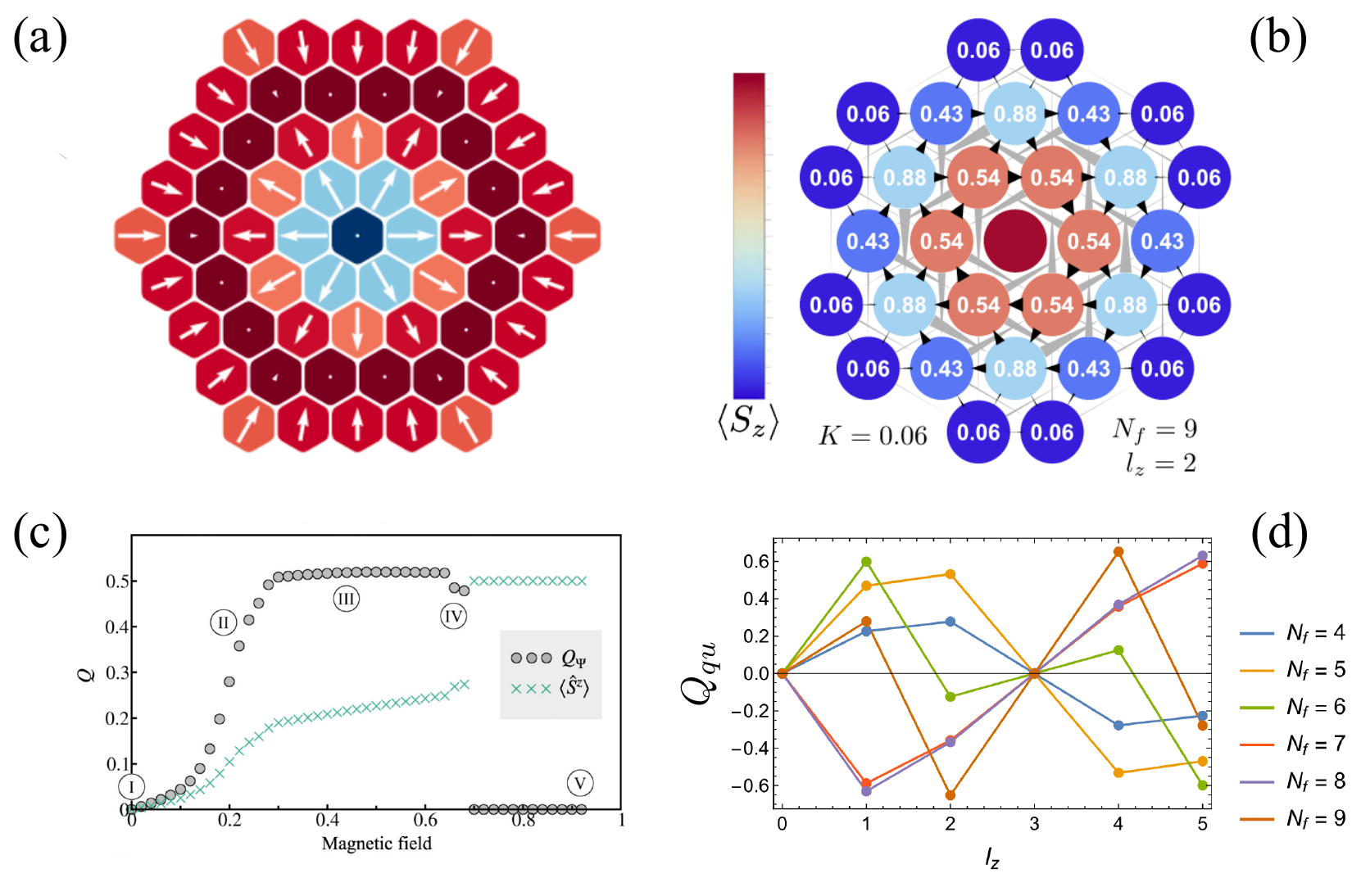}
	\caption{Quantum skyrmions. 
 (a) Local expectation value of the spin operator $\langle \vec S(\vec r_i) \rangle$ for a DMI system with $N=61$ lattice sites and open boundary conditions, that resembles a N\'eel skyrmion. The color coding represents the expectation value of the out-of-plane component $S_z(\vec r_i)$ \cite{Haller2022}. (b) Local expectation value for a system with frustrated interactions and $N=31$ using fixed boundary conditions, where the color coding represents $\langle S_z(\vec r_i) \rangle$. Here, the expectation value for the in-plane component vanishes, however, the quantum skyrmion possesses characteristic antiferromagnetic correlations as indicated by the numbers in the circles \cite{PhysRevX.9.041063}. (c) 
 The lattice averaged scalar spin chirality $Q_\Psi$ is an indicator of skyrmionic correlations in the ground state \cite{PhysRevB.103.L060404}. (d) Winding number of Eq.~\eqref{WindingNumber} evaluated for the frustrated system (b) with respect to eigenstates with quantum numbers of total spin $S_z = S_z^{\rm tot} - N_f$ and orbital angular momentum $l_z$ \cite{PhysRevX.9.041063}.}
 \label{Fig:Quantum_Skyrmion}
\end{figure*}

\subsection{Skyrmions in quantum magnetism} \label{Sec:SkyrmionQuantumMagnet}

The semiclassical treatments of the previous sections are only controlled for systems consisting of spins with large $S \gg 1$ or, in the continuum limit, with a large spin density $s$. For smaller $S$ like $S=1/2$ or $S=1$ it is more appropriate to treat spins from the very beginning as quantum operators instead of classical vectors. Using the spin-coherent states, approximate skyrmion wavefunctions have been proposed \cite{Schliemann1998,Istomin2000,Haller2024} that captured essential aspects of their semiclassical properties. Nevertheless, for quantitative studies, the corresponding spin Hamiltonian operator needs to be diagonalized with numerical methods, a common approach in quantum magnetism. As the Hilbert space increases exponentially with system size, numerical approaches are restricted to small systems. Exact diagonalization of the Hamiltonian, which yields all eigenvalues and eigenfunctions, is currently able to treat up to approximately 30 lattice sites $N$ for a spin $S=1/2$. 

The density-matrix renormalization group (DMRG) method is able to treat much larger systems of the order of several $100$ sites, but the standard implementation only gives access to the ground state. The small system sizes pose a challenge for the study of quantum skyrmions: the linear system size $L \sim \sqrt{N}$ of a 2D system needs to be sufficiently large to accommodate at least a single skyrmion with linear size $r_{\rm sk}$. 
To obtain a texture, $r_{\rm sk}$ should extend, at the same time, over several lattice sites with lattice constant $a$ leading to the constraint $1 \ll r_{\rm sk}/a < \sqrt{N}$. 
Several recent works studied single quantum skyrmions using exact diagonalization methods with $N = 19$ \cite{PhysRevB.103.L060404,Sotnikov2023},
$N=31$ \cite{PhysRevX.9.041063}, and $N=16$ \cite{PhysRevResearch.4.023111} for $S=1/2$ and $N=7$ for $S=3/2$ \cite{Gauyacq_2019}. 
Ref.~\cite{Haller2022} used DMRG for $S=1/2$ 
and reached even lattice sizes hosting several skyrmions. Most of these works considered skyrmions with DMI except Ref.~\cite{PhysRevX.9.041063} which studied skyrmions stabilized by frustration.  

How do we identify quantum skyrmions with the help of these numerical methods? As a first step, the local expectation value of the spin operator $\langle \vec S(\vec r_i) \rangle$ at lattice site $\vec r_i$ can be evaluated, e.g., with respect to the ground state, and compared with classical expectations. Here, the implemented boundary conditions are, however, instrumental. In case of periodic boundary conditions, this expectation value will be trivial $\langle \vec S(\vec r_i) \rangle =$ const.~because of translational invariance \cite{PhysRevB.103.L060404,Sotnikov2023}. 
Ref.~\cite{Haller2022} used instead open boundary conditions and the remaining works cited above implemented fixed boundary conditions by embedding the system in a ferromagnetically polarized background, both of which enforce a non-trivial profile on $\langle \vec S(\vec r_i) \rangle$. This allows to determine a quantum phase diagram for finite lattice sites $N$ for frustrated quantum skyrmions \cite{PhysRevX.9.041063} as well as for DMI stabilized quantum skyrmions \cite{Haller2022}, as shown in Fig.~\ref{Fig:PhaseDiagrams}-b) and c).

An example of the local expectation value $\langle \vec S(\vec r_i) \rangle$ for a system with N\'eel DMI is shown in Fig.~\ref{Fig:Quantum_Skyrmion}(a) reproducing 
the expected profile for a N\'eel skyrmion with a well-defined helicity. The local expectation value for a frustrated quantum skyrmion is shown in Fig.~\ref{Fig:Quantum_Skyrmion}(b). For an eigenstate, the expectation values for the in-plane components $S_x(\vec r_i)$, $S_y(\vec r_i)$ vanish and, consequently, also the helicity expectation value. However, there are characteristic antiferromagnetic correlations $C_\perp = -4 \langle S^x_i S^x_{\bar{i}}+ S^y_i S^y_{\bar{i}} \rangle$ between spins located on opposite sites $i$ and $\bar{i}$ with respect to the center as indicated by the numbers in the circles of Fig.~\ref{Fig:Quantum_Skyrmion}(b). These correlations are typical for skyrmionic configurations in a frustrated system.

The configuration space of classical spin textures, that are smooth on the length scale of the lattice spacing $a$, decomposes into topological sectors that are labeled by an integer winding number $Q$, see Eq.~\eqref{TopCharge}. Smooth fluctuations are not able to change the winding number, leading to the notion of topological protection. Does this concept still apply to quantum skyrmions? 
The definition of the winding number can be generalized to spins of size $S$ on a lattice 
\cite{BERG1981412,4121581,PhysRevX.9.041063}
\begin{align}
\label{WindingNumber} 
&Q_{qu} = \\ \nonumber
&\frac{1}{2\pi}\sum_{\langle ijk\rangle} \arctan \frac{\frac{1}{S^3}\langle \mathbf{\hat{S}}_i\cdot (\mathbf{\hat{S}}_j \times \mathbf{\hat{S}}_k) \rangle}{1+\frac{1}{S^2}(\langle \mathbf{\hat{S}}_i \cdot \mathbf{\hat{S}}_j \rangle + \langle \mathbf{\hat{S}}_i \cdot \mathbf{\hat{S}}_k \rangle +\langle \mathbf{\hat{S}}_j \cdot \mathbf{\hat{S}}_k \rangle)}
\end{align}
where the sum extends over a triangulation of the lattice. In the classical limit, $Q_{qu}$ is an integer. If the expectation values in Eq.~\eqref{WindingNumber} are evaluated with respect to a quantum state, this is in general not the case. In Fig.~\ref{Fig:Quantum_Skyrmion}(d) the value for $Q_{qu}$ is shown evaluated for various eigenstates of the frustrated system labeled by quantum numbers of the total $S_z = S^{\rm max}_z - N_f$ and the orbital angular momentum $l_z$. From analogy with the classical solution, it is expected that the quantum skyrmion and antiskyrmion is an eigenstate with eigenvalues $S_z \pm l_z$, respectively. This leads to selection rules that explain the location of maxima and minima of the winding number as a function of $l_z$ and $N_f$. In particular, for the hexagonal flake of Fig.~\ref{Fig:Quantum_Skyrmion}(b) the winding number vanishes for eigenstates with $l_z = 0$ and $l_z = 3$ indicating that they are superpositions with equal skyrmionic and antiskyrmionic content \cite{PhysRevX.9.041063}.
It is an open question how these properties evolve with increasing system size $N$ for 
quantum skyrmions extending over more lattice sites. It remains to be studied whether the extremal values of $W$ increase further approaching an integer for $r_{\rm sk}/a \to \infty$ and whether in this limit an approximate topological protection emerges even for quantum skyrmions.

Another important quantity characterizing noncoplanar magnetic order is the scalar spin chirality $\chi_{ijk} = \mathbf{\hat{S}}_i\cdot (\mathbf{\hat{S}}_j \times \mathbf{\hat{S}}_k)$ for spins on a lattice. In the classical continuum limit, $\chi_{ijk}$ reduces to the topological density \eqref{TopChargeDensity} and the lattice sum $Q_\Psi = \frac{1}{\pi S^3} \sum_{\langle ijk \rangle} \langle \chi_{ijk} \rangle$ approximates the winding number. Thus, the expression $Q_\Psi$ should also be a good indicator for quantum skyrmions. The evolution of $Q_\Psi$ as a function of magnetic field for a small system $N=19$ with DMI is shown in Fig.~\ref{Fig:Quantum_Skyrmion}(c). It depicts a robust plateau in the field range where skyrmionic states are expected classically \cite{PhysRevB.103.L060404}.

\section{Accessing Quantum Aspects of Skyrmions in the Laboratory}

\subsection{Materials and architectures}

\subsubsection{Skyrmions capable of displaying quantum properties}\label{SubSec:Materials}
\begin{figure}[htbp]
\includegraphics[clip=true, width=0.99\columnwidth]{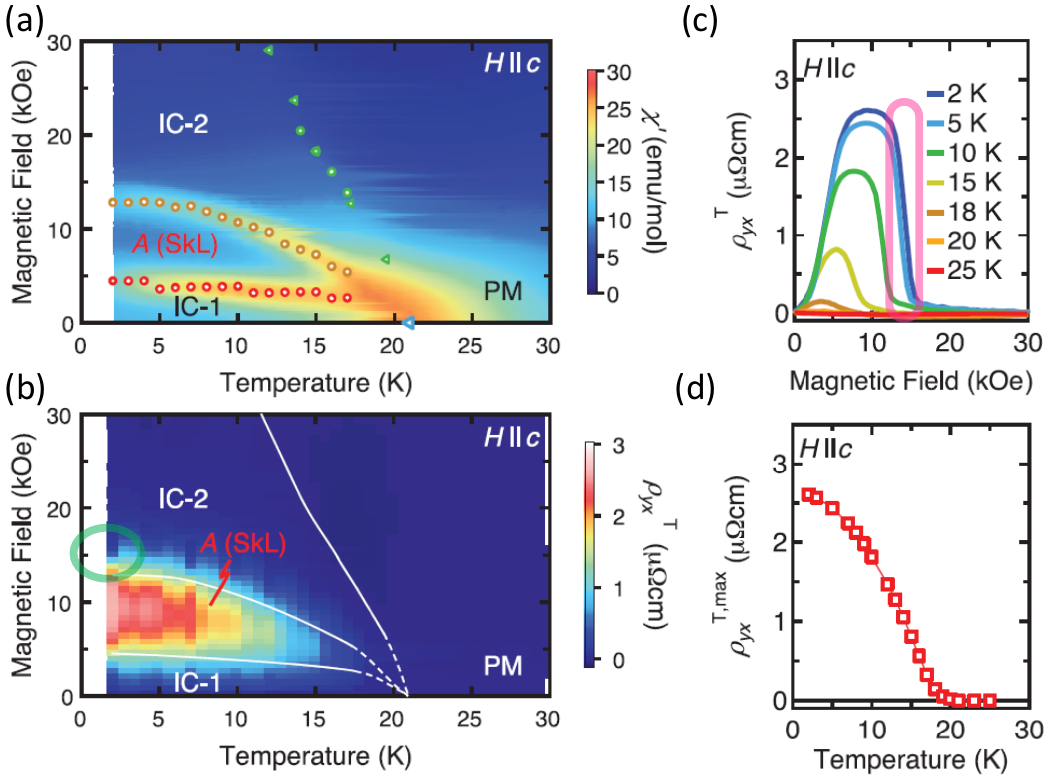}
\caption{\label{Fig7} Targeting isolated skyrmions in Gd$_2$PdSi$_3$. (a) Magnetic susceptibility of Gd$_2$PdSi$_3$ in the $H,T$ plane, illustrating the various non-collinear phases. (b) Evolution of the topological Hall resistivity in the same parameter range, which peaks in the skyrmion lattice phase identified by resonant x-ray scattering. A green circle highlights the target zone for finding isolated quantum skyrmions. (c) Field dependence of $\rho_{yx}^T$: a ``tail'' emerges at the transition from skyrmion lattice to the trivial high-field phase, highlighted by the pink oval. (d) Temperature dependence of the maximum $\rho_{yx}^T$, implying no loss of topological charge or scalar spin chirality in the $T\rightarrow0$ limit. Data adapted from ref.~\cite{Kurumaji2019}.}
\end{figure}

From the recently identified centrosymmetric skyrmion materials~\cite{Hirschberger2019,Khanh2020,Takagi2022}, Gd$_2$PdSi$_3$ stands out as the most plausible source of quantum skyrmions~\cite{Kurumaji2019}. This compound features strongly frustrated magnetism due to its hexagonal symmetry and nesting-enhanced RKKY interaction~\cite{Inosov2009}, resulting in a 2.4\,nm period skyrmion lattice persisting down to zero temperature (unlike the thermally-stabilized skyrmions in Gd$_3$Ru$_4$Al$_{12}$).  The magnetic susceptibility and topological Hall resistivity are shown in Fig.~\ref{Fig7}: a tail in the topological Hall signal at the upper boundary of the skyrmion lattice phase may be a signature of isolated skyrmions rather than a dense lattice. Spatially-resolved experiments are required to verify the local magnetic structure in this field range. 

Experimental confirmation of nanoscale skyrmion crystals in frustrated magnets may allow the detection of novel phenomena including the quantized topological Hall effect~\cite{Hamamoto2015}, or topological magnon insulators with chiral magnonic edge modes~\cite{Roldan-Molina2016,Diaz2020}. However, skyrmion lattices will be difficult to re-purpose for device applications.  So far, \emph{isolated} skyrmions have not been experimentally reported in frustrated magnets, despite predictions for their stability in a triangular lattice with ferromagnetic nearest-neighbor and antiferromagnetic next-nearest-neighbor interactions~\cite{Leonov2015}. Identifying materials whose parameters match this model is a priority: the oscillatory nature of the RKKY interaction suggests that metallic hexagonal magnets 
are promising candidates.  

However, an ultra-low Gilbert damping ($\alpha\lesssim10^{-4}$) will be essential for skyrmion qubits to achieve microsecond coherence times competitive with superconducting qubit architectures. Since magnon scattering off itinerant electrons is a major contributor to damping, it would be preferable to avoid metallic skyrmion hosts for qubit applications: this would eliminate RKKY-mediated stabilization and considerably reduce the pool of candidate materials. Meanwhile, one could seek a semimetallic or doped semiconducting frustrated magnet with a low density of states~\cite{Schoen2016}.  Semiconducting ternary manganese chalcogenides may be a fruitful area to explore since the strongly frustrated Mn triangles in their crystal structure develop non-collinear double-$q$ ordering at low temperature~\cite{Pak2019}. 

Individual skyrmions could also be isolated by geometric confinement in magnetic nanodot arrays. This is an appealing concept since entanglement between quantized skyrmions (mediated via dipolar interactions) could be adjusted via the array period. However, fabricating nanodots sufficiently small for their skyrmions to exhibit quantization ($r_{\mathrm{sk}}\lesssim5$\,nm) will be more lithographically demanding than isolating DMI skyrmions~\cite{Ho2019}. Edge and demagnetization effects may also influence skyrmion dynamics: these factors await exploration in frustrated centrosymmetric systems.

Entering the quantum regime also raises the prospect of a novel stabilization mechanism: could quantum fluctuations replace the thermal fluctuations that sustain skyrmions in the B20 compounds~\cite{doi:10.1126/science.1166767,Belitz2018}? Initial indications are positive: the zero-point (vacuum) energy associated with quantum spin fluctuations is predicted to generate an effective Casimir field which increases the skyrmion stability, raising the annihilation field by $\sim$\,10\%~\cite{Roldan-Molina2015}. Analogous to thermal fluctuations, quantum fluctuations should also increase topological charge density~\cite{Keesman2015a}, thus generating large emergent topological Hall signals at skyrmion phase boundaries even in the $T\rightarrow0$ limit.  This effect may explain the discrepancy between the field range in which the topological Hall amplitude is maximal~\cite{Shang2021} and the phases in which low-temperature neutron scattering reveals skyrmions~\cite{Takagi2022} in EuAl$_4$.

\subsubsection{Skyrmion platforms for topological superconductivity}
The minimal requirement for a skyrmion to induce a local topological phase transition in a neighboring superconductor is for the $k=0$ gap opened by its Zeeman field, $\Delta_Z \equiv g^*\mu_B B_Z/2$, to exceed the pairing gap $\Delta_0$.  In a chiral multilayer comprising $3d$ magnetic elements, stray field-coupled to a Nb thin film, the flux density above the skyrmion cores cannot exceed the saturation magnetization $M_s \lesssim 2$\,T. Assuming an effective $g$-factor $g^*=2$, we obtain $\Delta_Z \sim 0.1$\,meV\,$\ll\Delta_0\sim1$\,meV and hence the superconductor will remain topologically trivial. One solution is to devise a multilayer in which superconductivity is induced in a semiconductor with a large $g^*\gg20$, e.g. (Cd,Mn)Te~\cite{Fatin2016}, capped by a skyrmion-hosting material. This $g$-factor constraint vanishes for exchange-coupled superconductor-skyrmion hybrids, since typical exchange fields in metallic ferromagnets $h_{\mathrm{ex}}\gtrsim$~100\,T. These strong fields may persist up to a distance of $\xi$ into the superconductor due to the inverse proximity effect~\cite{Bergeret2004}, rendering a topological phase transition more plausible.  

Exchange-coupled superconductor/non-collinear magnet systems still pose challenges for generating and manipulating MZM.  Strong SOC from the Rashba effect and/or heavy elements at the interface (e.g. Pt monolayers) will be essential for individual skyrmions to nucleate vortices.  Also, the electrical contact between magnet and superconductor implies that skyrmion-vortex pairs cannot be moved via spin-orbit or spin-transfer torques as previously envisaged~\cite{Nothhelfer2022,Konakanchi2023}, since applied currents will be shorted through the superconductor.  These constraints suggest that the most immediate route towards MZM braiding via hybridized topological solitons will rely on non-contact techniques, such as coupling between skyrmions and local field gradients from atomically-sharp magnetic scanning probes~\cite{Casiraghi2019}.

Achieving strong skyrmion-vortex interactions \emph{and} a robust topological phase transition therefore requires exchange coupling between magnet and superconductor, as well as a high tunability of the skyrmion radii and densities (to optimally match the superconducting lengthscales). Multilayer stacks based on elemental magnets and strong SOC metals - e.g. Ru/Co/Pt~\cite{Palermo2020} and Ir/Fe/Co/Pt~\cite{Soumyanarayanan2017} offer precise control over the skyrmion morphology by engineering the stack geometry. So far, these materials have been grown by magnetron sputtering, resulting in granular or amorphous structures with inhomogeneous magnetic properties~\cite{Bacani2019} and high critical current densities for skyrmion motion~\cite{Tan2021a}. The next step will be to reduce heterogeneity in these architectures using lower-energy deposition techniques: recent efforts via e-beam evaporation appear promising~\cite{Bromley2022}. An alternative possibility would be to synthesize non-centrosymmetric PtMnGa in thin film format, capped with an ultra-thin layer of Nb. Lamellae cut from PtMnGa crystals by a focused ion beam develop N\'eel skyrmions whose radii evolve from $\lesssim$\,50\,nm to $\sim$\,350nm upon increasing the lamellar thickness~\cite{Srivastava2020}. Crucially, the skyrmions remain stable at 5\,K, thus offering a low-disorder environment in which to explore skyrmion-vortex coupling. 

\subsection{Validating quantum properties through advanced experimental methods}
To explore the potential for using skyrmions in future quantum operations, it is imperative to confirm their theoretically-predicted properties and behavior \cite{Gobel2021}. Advanced techniques with atomic-scale spatial resolutions are necessary to image nanoscale skyrmions, while temporal resolutions down to the femtosecond regime will be required to facilitate studies of skyrmion dynamics in monophase materials as well as devices. Hybrid architectures will benefit from experimental approaches with high levels of layer and chemical sensitivities. Measurements of the quantum states of skyrmions will be accomplished via high-resolution spectroscopic techniques, such as angle-resolved photoemission spectroscopy (ARPES) or scanning tunneling spectroscopy (STS). {\em In-situ} characterizations at ultra-low temperatures and in variable magnetic fields will be critical to explore quantum-enabled functionalities, as will the capability to apply electrical currents {\em in-operando}. 

\begin{figure}[htbp]
\includegraphics[clip=true, width=0.8\columnwidth]{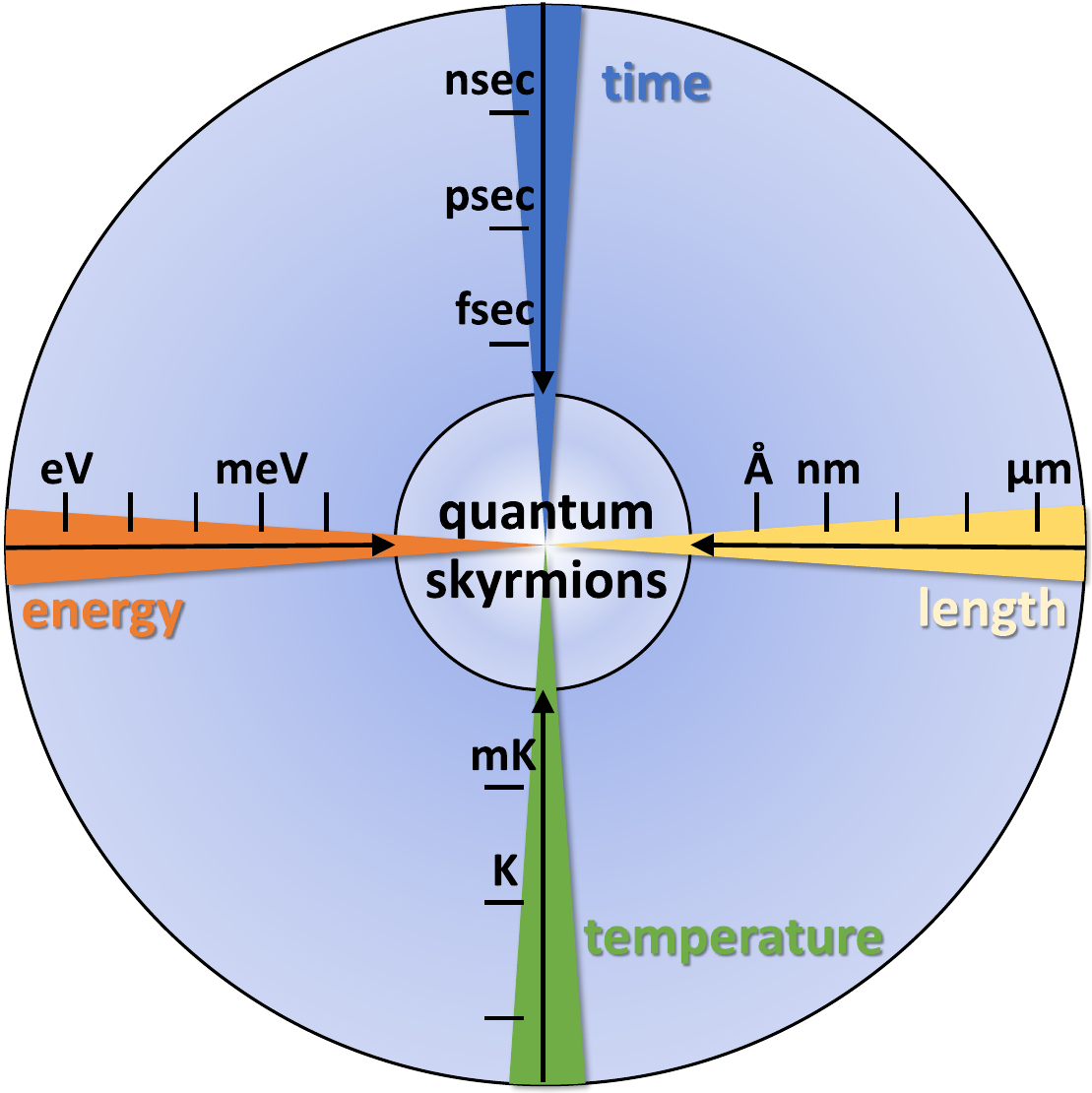}
\caption{\label{Fig9} Experimental validation of quantum properties of skyrmions require progress along four primary axes: length, time, energy, and temperature. Progress will be enabled by (i) increasing energy resolution (e.g. to expose skyrmion quantization), (ii) decreasing experimental temperatures (hence suppressing thermal smearing), (iii) increasing spatial sensitivity to resolve individual nanoskyrmions and (iv) decreasing experimental timescales to access quantum fluctuations.}
\end{figure}

The experimental approaches can be broadly divided into two categories. The first category measures global (macroscopic) spatially and temporally averaged quantities, e.g. bulk electrical transport (see \ref{transport}), or the total magnetization of an \emph{ensemble} of quantum skyrmions. These methods also include techniques capable of probing average quantities at {\AA}ngstr{\"o}m lengthscales and/or with atomic specificity, e.g. advanced neutron or x-ray diffraction techniques.

The second category directly characterizes the local (microscopic) properties and behavior of \emph{individual} quantum skyrmions with spatial resolution. These include the most advanced new imaging techniques harnessing the short wavelengths and coherence of x-rays (see \ref{xray}), such as holography and ptychography. Scanning transmission microscopies using electrons (STEM) or x-rays also offer high local sensitivities (see \ref{tem}), as do scanning probe techniques (see \ref{SPM}) such as nitrogen-vacancy (NV) microscopy or scanning tunneling microscopy (STM). 

Fig.~\ref{Fig9} illustrates schematically the experimental advances that will be critical for studying quantum skyrmions, especially the phenomena or processes outlined in this Colloquium. Our target of exposing the quantum properties of skyrmions requires us to improve experimental efforts along four key experimental axes: length, time, energy, and temperature. It is unlikely that a single experimental technique will be able to deliver on all four axes, but a common characteristic of the most sophisticated methods is that they provide excellent performance in two or more of these four directions. For instance, x-ray spectroscopy combined with atomic spatial resolution in soft x-ray-STM or nano-ARPES will enable element-specific visualization at length scales that are highly relevant to quantum phenomena. Femtosecond time-resolved measurements at the nanoscale in x-ray photo correlation spectroscopy or transmission electron microscopy provides another example: these ultra-fast techniques will deliver new insights into atomic-scale fluctuations in quantum skyrmions. In the following we describe briefly a few selected examples of ongoing research, which we believe have the potential to open new avenues that are most relevant for the validation of quantum skyrmions.

\subsubsection{\label{tem}Advances in electron-based techniques}
The recent push towards operating a variety of electron microscopes at ultralow temperatures ($<$\,1K) has opened an important path towards studying quantum materials exhibiting low-temperature phase transitions, as well as superconducting quantum information systems \cite{McComb2019, Minor2019}. Spin-polarized low energy electron microscopy (SPLEEM) has already enabled accurate mapping of spin chiralities in multilayers at room temperature \cite{ChenNatComm2013, ChenNatComm2017}. Such measurements can now be extended to the cryogenic regime, via quantum SPLEEM (which can reach $\sim 4$\,K) and 4D-STEM (which records 2D electron diffraction patterns across a 2D grid in real space and has previously revealed the local strain state with sub-nm resolution in a polar skyrmion system~\cite{ShaoNatureComm2023}). High resolution electron tomography has reached a level to  locate every atom with high precision in a nanoparticle consisting of thousands of atoms~\cite{RN72}. These methods offer sufficient resolution to map the local polarization and chirality of individual quantum skyrmions.

The dynamics of quantum skyrmions can be explored by adding temporal resolution to ultrafast electron microscopy: a new technique relying on pulsed-laser photoemission to generate electron bunches in a transmission electron microscope (TEM) column. Pump-probe studies in which two sequential laser pulses are coupled into a TEM column have already enabled the exploration of sub-microsecond to sub-picosecond dynamics in magnetic vortices \cite{SilvaPRX2018, JingUltramicroscopy2019, PhysRevResearch.4.013027}.

\subsubsection{\label{xray}Advances in x-ray and UV spectromicroscopies}

A rapidly emerging area of research exploits electromagnetic beams with large orbital angular momentum (OAM). These have been demonstrated in TEM \cite{McMorranScience2011}, where electron beams carrying quantized OAM up to $100\hbar$ per electron were observed. In modern optics, ``twisted light'' carrying OAM provides an additional degree of freedom and is an emerging resource for both classical and quantum information technologies \cite{PhysRevLett.121.233602}. The high x-ray coherence available at 4th generation x-ray facilities (such as x-ray free electron lasers (XFEL) or Diffraction Limited Storage Rings (DLSR)) now enables the analogous creation of x-ray vortex beams with large OAM \cite{LeeNatPhotonics2019}. These beams will be capable of directly probing highly excited states in quantum skyrmions. 

The full coherence of 4th generation light sources is also facilitating unprecedented magnetic x-ray microscopy techniques. X-ray laminography \cite{WitteNanoLett2020} enables spatially resolved 3D nanotomography of spin textures, while quantitative analysis of the x-ray magnetic circular dichroism (XMCD) contrast represents a first step towards quantitative magnetic nanometrology \cite{raftrey2023quantifying}. The recent addition of time resolution to x-ray laminography has created the ultimate tool for imaging spatio-temporal behavior of nanoscale spin textures \cite{DonnellyNatNano2020}.  
 
Nano-ARPES can resolve the electronic structure of materials as a function of position, momentum and (in principle) spin \cite{XieChemMat2023}. This technique has the potential to study quantum skyrmions at upcoming DLSR sources with full x-ray coherence \cite{RotenbergJSR2014}. Momentum microscopes take a step further \cite{KarniAdvMat2023}, combining ARPES with a photoemission electron microscope to provide lateral information with spectroscopic analysis. More recently, time-resolved momentum microscopy \cite{Fedchenko2024} has enabled studies of the nature and dynamics of occupied excited states. 
 
One can also combine the chemical, magnetic, and structural sensitivity of x-rays with the ability of STM to resolve and manipulate atoms on surfaces. Following this approach, quantum tunneling processes have been revealed via x-ray-excited resonance tunneling in individual atoms \cite{AjayiNature2023}. This opens a path towards the simultaneous ({\em multimodal}) characterization of elemental and chemical properties of quantum skyrmions at the ultimate single-atom limit. 
 
New fully coherent XFEL facilities offer a novel approach to study ultrafast dynamical processes. Ultrashort x-ray pulses with high intensity, tunable photon energy, and polarization have demonstrated the capability to obtain femtosecond (fs) single-shot images of nanoscale magnetic order using x-ray holography \cite{WangPRL2012}. Implementing a two-pulse operational mode into the XFEL beam - where two short (fs) consecutive pulses interact with spin textures in the sample, separated by variable time delays in the pico to nanosecond regime - has demonstrated the ability to probe nanoscale spin fluctuations \cite{AssefaRSI2022}. Recent experiments depicted a sharp difference in the characteristic timescales of such nanoscale spin fluctuations in an amorphous GdFe multilayer, depending on whether the system orders in a hexagonal skyrmion lattice or a stripe phase \cite{SeabergPRL2017}.

X-ray ferromagnetic resonance (X-FMR) combines two distinct techniques, namely XMCD or x-ray magnetic linear dichroism  and ferromagnetic resonance. It has been used to explore aspects of magnetization dynamics, such as spin currents in magnetic multilayers \cite{vdLaanNIM2023}. The ability to obtain element and phase resolution enables detailed investigations of spin current propagation mechanisms, while the high sensitivity of X-FMR allows the detection of ultrasmall signals. In addition to time-resolved information, reflection geometry X-FMR measurements enable depth-resolution of the complex magnetization dynamics in a wide range of functional heterostructures \cite{PhysRevLett.125.137201}. Combining microwave pump pulses in synchrony with the ultrafast x-ray pulses of an XFEL, one can envision probing Rabi oscillations in quantum skyrmions. 

\subsubsection{\label{transport}Transport and microwave spectroscopies}  
Aside from scattering techniques, experimental inference of dense nanoskyrmion lattices has generally been restricted to observations of the topological Hall resistivity $\rho_{yx}^T(H)$~\cite{Kurumaji2019,Hirschberger2019,Shang2021} and the related topological Nernst effect~\cite{Hirschberger2019a}. Accurate determination of $\rho_{yx}^T(H)$ at temperatures below 1.8\,K is a challenge, due to the scarcity of dc magnetometers operating in the milliKelvin regime. The difficulties of signal compensation in second-order gradiometers (previously solved by vibrating the entire refrigerator around samples~\cite{Morello2005}) could be tackled by modern low-dissipation piezo manipulators, making the development of a ``plug 'n' play'' SQUID magnetometer for modern dry dilution refrigerators a realistic target. Even if this technical barrier is overcome, a non-zero $\rho_{yx}^T$ is not exclusively sensitive to skyrmions~\cite{Kimbell2022} and is generated by any static or fluctuating spin configuration with non-zero scalar spin chirality $\mathbf{S_i} \cdot (\mathbf{S_j} \times \mathbf{S_k})$~\cite{Wang2019,Fujishiro2021,Raju2021}. At best, $\rho_{yx}^T$ is an indicator of a finite real-space Berry phase and does not provide information on the nanoscopic spin structure or the absolute magnetic parameters. 

These deficits can be largely eliminated by a complementary technique: broadband coplanar waveguide (CPW) FMR spectroscopy. In non-collinear magnets, the spin dynamics develop multiple resonances at microwave frequencies~\cite{Schwarze2015}, whose hierarchy and dispersions are acutely sensitive to the density and helicity of skyrmions~\cite{Garst2017}. Microwave absorption studies of centrosymmetric skyrmion materials therefore offer a suitable method of confirming the proposed dipolar interaction-driven Bloch helicity, as well as verifying the predicted low-frequency helicity rotational mode (Sec.~\ref{Sec:Tunable_Helicity}). 

In putative nanoscale skyrmion hosts unexplored by x-ray scattering, one can vary the orientation of the microwave magnetic field $h_{RF}$ to selectively enhance skyrmion eigenmode excitations, hence identifying their helicity. Out-of-plane $h_{RF}$ (applied by centering the sample over a loop-shaped CPW) preferentially excites skyrmion breathing modes, whereas in-plane microwave fields induce CCW skyrmion gyration. The breathing frequency exceeds the lowest CCW mode in skyrmions with Bloch helicity but falls below the CCW mode in N\'eel skyrmions~\cite{Mruczkiewicz2017,Zhang2017f}. Topological charge reversal to an antiskyrmion state reverses the circulation of the lowest energy gyrational mode~\cite{Liu2020f}, causing another shift in the resonance spectra. 

Advances in cryogenic microwave hardware now permit such experiments at low incident microwave power below 1\,K. Combining cryogenic attenuators on the input line with circulators and a high electron mobility transistor amplifier at 4\,K on the output has enabled resonance measurements in yttrium iron garnet at signal powers approaching -100\,dBm and temperatures down to 20\,mK~\cite{VanLoo2018,Kosen2019a,Wolz2021}. Operation with Josephson or traveling-wave parametric amplifiers would permit experiments with lower photon counts (i.e. reduced sample heating), at the cost of a smaller bandwidth. 

The interaction of magnetic vortices, domain walls, or skyrmions with spin waves (magnons) offers opportunities to design active components for data storage and signal processing \cite{BarmanJPhys2021}. Skyrmion arrays are particularly well-suited for future magnonic devices operating at microwave frequencies, due to their high degree of controllable reconfigurability. Time-resolved studies of magnetic vortex dynamics have demonstrated the feasibility of generating propagating short-wavelength magnons \cite{WintzNatNano2016}, while spin wave emission in the Ku frequency band has been observed from skyrmionic bubbles \cite{Srivastava2023}. 

\begin{figure}[t]
\includegraphics[clip=true, width=0.99\columnwidth]{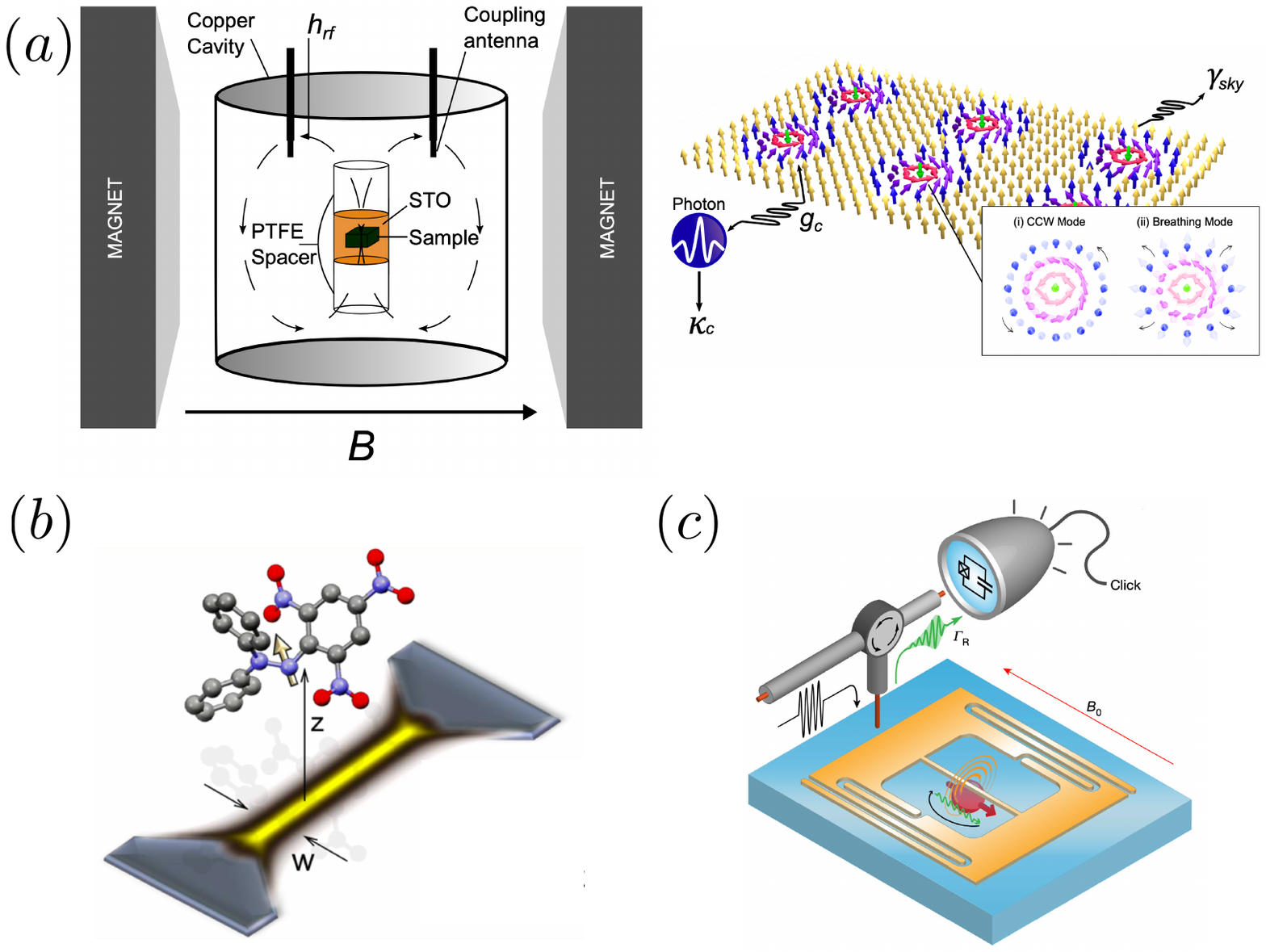}
\caption{Microwave resonators can develop strong interactions between their cavity modes and skyrmion excitations. a) Schematic view of a 3D cavity hosting a sample of Cu$_2$OSeO$_3$. The system exhibits a high cooperativity for the coupling of cavity photons to the CCW and breathing skyrmion lattice modes. Figure adapted from Ref.~\cite{Khanh2020}. In the quantum limit of a few spins, significant coupling factors have been demonstrated by incorporating nanoscopic constrictions in the central transmission line of (b) superconducting coplanar resonators coupled to molecular spins and (c) planar resonators coupled to single spins. Figures adapted from Refs. \cite{Wang2023} and \cite{Gimeno2020}.} \label{Fig:Cavities}
\end{figure}

The coherent coupling $g$ between magnetic systems and microwave photons provides an appealing method of magnetic mode detection. On a classical level, ferromagnetic systems have been successfully coupled to microwave resonators~\cite{Huebl2013,Li2019b,Hou2019,Ghirri2023,Gonzalez2024}. Additionally, interactions between the microwave modes of 3D cavities and the gyrational and breathing modes of skyrmion lattices have been demonstrated in bulk Cu$_2$OSeO$_3$ crystals~\cite{Khan2021,PhysRevB.104.L100415} [Fig.~\ref{Fig:Cavities} (a)]. This system exhibits a high cooperativity, making it a promising platform for exploring light-matter interactions. In the strong coupling regime, the energy levels of the combined system form new hybrid states, which - in the low loss and small cavity linewidth limits - would manifest as two distinct peaks in the resonator transmission spectrum, with a frequency splitting proportional to $g$. The strong coupling displayed by the skyrmion lattice phase in bulk Cu$_2$OSeO$_3$ is believed to originate from the high topological charge density, thus suggesting that even stronger interactions may be achievable for dense nanoskyrmion arrays.

In the quantum limit, where small skyrmions consist of only a few spins, small cavities offer a significant advantage as $g$ increases with decreasing mode volume. This coupling can be further enhanced by using superconducting resonators, which offer improved quality factors. The capability of modern circuit quantum electrodynamic hardware to study and control light-matter interaction at the quantum level \cite{RevModPhys.93.025005} hence offers a realistic path towards experimentally demonstrating helicity quantization in nanoskyrmions. This is further supported by the elevated single spin-photon coupling factors that have already been achieved by incorporating nanoscopic constrictions in the central transmission line of superconducting coplanar resonators \cite{Wang2023,Gimeno2020} [Fig.~~\ref{Fig:Cavities} (b),(c)]. Outside the strong coupling regime, energy level quantization can also be detected in the dispersive limit through frequency shifts, which reveal the underlying quantum properties of the system \cite{RevModPhys.93.025005}.

\subsubsection{\label{SPM}Scanning probes adapted to skyrmions and superconductor-skyrmion hybrids}  

Nitrogen-vacancy (NV) center microscopy is useful to analyze nanoscale non-collinear spin textures since it has no perturbative effect on the sample magnetization and directly measures the absolute stray field. The flux sensitivity lies in the low pT/$\sqrt{\mathrm{Hz}}$ range for vacancy ensembles~\cite{Barry2020} and can reach 100\,nT/$\sqrt{\mathrm{Hz}}$ for single-vacancy scanning gradiometric probes~\cite{Huxter2022}, which should be sufficient to infer the helicity of a static skyrmion. 

Spatial resolution of a diamond NV-center probe is controlled by the sample-vacancy separation, which approaches 10\,nm via controlled vacancy implantation in lithographically-defined tips.  This performance is confirmed by successful helicity imaging in skyrmion microdots~\cite{Dovzhenko2018} and vorticity detection in antiferromagnets~\cite{Tan2024}. However, NV-center microscopy has not been demonstrated below 350\,mK due to heating from the laser excitation and microwave drive ~\cite{Scheidegger2022}. Progress in coupling diamond nanobeams to tapered optic fibers~\cite{Li2023b} may eventually breach this limit, allowing NV sensors to operate in a regime where skyrmion quantization could be discernible.

Spin-polarized STS offers atomic-level spin sensitivity while imaging non-collinear magnets in milliKelvin environments~\cite{Heinze2011,Herve2018}. However, it is restricted to metallic systems and requires separation of the topographic, electronic, and magnetic contributions to the tunneling spectra. The polarization of ferromagnetic tips becomes fragile in the low-moment (non-perturbative) limit, although this can be mitigated by Cr-based antiferromagnetic tips. Dynamic resolution is achievable via homodyne detection of a dc offset in the spin-polarized tunnel current, which emerges due to the precession of spin textures resonating in an external microwave field~\cite{Herve2019}. 

The sensitivity of STS to the local density of states may reveal the impact of proximate magnetism on superconductivity in skyrmion-vortex systems.  However, unless the surface of the superconductor is fully spin-polarized by the underlying spin textures, STS is blind to the underlying magnetization and cannot distinguish vortex polarities. Also, the anticipated topological $p_x \pm ip_y$ order parameter is fully gapped and difficult to distinguish from trivial $s$-wave pairing, while the pitfalls of MZM detection in vortex cores are well-documented~\cite{Jack2021}.  Josephson STS might be a more suitable tool for imaging emergent $p$-wave pairing in superconductor-skyrmion hybrids, with the proviso that the tip must possess a similar odd-parity gap function~\cite{Mascot2021}. 

Direct imaging of skyrmion-vortex pairs via their stray magnetic fields is challenging since the flux density associated with the Pearl vortex supercurrents is several orders of magnitude smaller than the skyrmion stray field [Fig.~\ref{Fig6}(d)].  Here, \emph{indirect} pair imaging with magnetic force microscopy is viable by measuring the expected change in the skyrmion radius upon cooling through $T_c$ (in response to the stray field from the incipient (anti)-vortex)~\cite{Andriyakhina2021b,Andriyakhina2022,Apostoloff2023,Apostoloff2024}.  

The morphology of the skyrmion-vortex pair is controlled by the skyrmion helicity, radial spin profile, and skyrmion/vortex stacking order, with particularly complex behavior in N\'eel skyrmions due to their asymmetric stray field profile~\cite{Dahir2020}.  In certain configurations, the equilibrium vortex position may be laterally displaced from the skyrmion, which can be qualitatively understood as an alignment of the in-plane stray-field components to minimize the free energy. Non-uniform magnetic fields generated by vortices may also permit stabilization of high-$Q$ skyrmions~\cite{Shustin2023}.  More pertinently for quantum applications, this non-uniformity could provide the local in-plane field gradient necessary to break the helicity degeneracy in a skyrmion qubit~\cite{PhysRevLett.127.067201}.

\begin{figure}[htbp]
\includegraphics[clip=true, width=0.99\columnwidth]{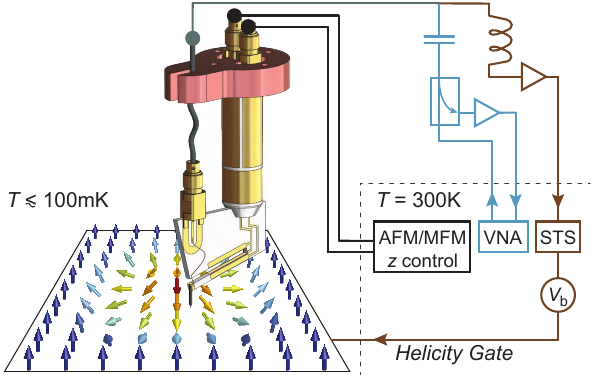}
\caption{\label{Fig10} Hybrid scanning probe microscopy combining multiple independent detection circuits. Self-sensing quartz tuning forks are sufficiently stiff to permit tunneling measurements as well as non-contact force microscopy and are compatible with milliKelvin environments. A bias tee in the tip coaxial line separates low-frequency from high-frequency signals: connecting a Vector Network Analyzer (VNA) to the high-frequency branch thus enables local microwave absorption spectroscopy at arbitrary dc bias voltages $V_\mathrm{b}$. Deposition of a low-moment magnetic coating on the tip adds another layer of sensitivity, allowing the instrument to perform Atomic/Magnetic Force Microscopy (AFM/MFM) and spin-polarized STS.}
\end{figure}

A new local probe variant may offer the ideal combination of spectroscopic resolution and helicity sensitivity to explore superconducting hybrids as well as nanoskyrmions in the quantum limit.  Near-field microwave microscopy measures the complex reflection coefficient between a sharp tip and sample at GHz frequencies~\cite{Barber2022}. Spatial resolution below 5\,nm has been reported~\cite{Lee2020}, as well as operation below 100\,mK~\cite{Jiang2023a,Cao2023} with single-photon sensitivity~\cite{Geaney2019}. 

Beyond detecting the local conductivity and carrier density, the high bandwidth allows scanning microwave microscopes to function as local magnetic resonance probes with spatially-resolved sensitivity to spin dynamics~\cite{Berweger2022}. The conducting tip could be used as a local gate to tune the $V_2$ component of the helicity potential (Sec.~\ref{Sec:MQT_Helicity}) while simultaneously probing the microwave absorption spectrum of an underlying skyrmion. With the addition of a bias tee to filter off low-frequency current (Fig.~\ref{Fig10}), a tuning fork microwave microscope could also perform conventional tunneling spectroscopy, analogous to a combined atomic force/scanning tunneling microscope~\cite{Coissard2023}. 

A key technical barrier to such measurements lies in the impedance-matching networks coupling the 50\,$\Omega$ transmission line to the (higher-impedance) tip-sample junction, which currently operates at discrete frequencies separated by $\geq$\,500\, MHz. Development of a \emph{continuous-frequency} broadband microwave microscope would enable local characterization of skyrmion eigenmodes in nanoscale geometries for both insulating and metallic magnets,  chiral Majorana edge state detection~\cite{Kamra2024} in skyrmion-mediated topological superconductors, as well as a dispersive readout of the MZM parity~\cite{Ren2023} via microwave spectroscopy of two fused skyrmion-vortex pairs: a unique suite of abilities. 

\subsection{Skyrmion devices for quantum technologies} \label{SubSec:Devices}

Microwave signals are a logical choice for probing and manipulating quantum skyrmions, given their quantization energy scales ($\sim$~100\,$\mu$eV) and coherence times ($\sim$~1\,$\mu$s). Here we may greatly benefit from recent technical and protocol advances by the quantum computing community.  Non-destructive readout of a solid-state qubit typically involves dispersive coupling of the qubit to a photonic cavity mode, such as a coplanar microwave resonator~\cite{Blais2004}. For spin qubits in semiconductors, this coupling occurs between the strong electric field in a superconducting resonator and the electric dipole moment of the qubit (which hybridizes with its spin degree of freedom in the presence of a magnetic field gradient)~\cite{Mi2018a,Landig2018,Samkharadze2018,DAnjou2019}. 

Any skyrmion in a centrosymmetric insulating magnet with helicity $\varphi_0\neq\pm\pi/2$ will also develop an electric dipole moment via the magnetoelectric effect, whose amplitude is proportional to $\cos \varphi_0$. Coupling individual skyrmion-helicity qubits or skyrmion arrays to a lithographically defined resonator will introduce a state-dependent phase shift into the microwave reflection coefficient, thus enabling readout within a framework compatible with existing quantum information hardware. Readout of a topological qubit based on skyrmion vortex pairs can also be accomplished via microwave techniques. The two states of a topological qubit correspond to the charge occupancy of a fused pair of MZM.  Although both states are degenerate while the MZM are distant, fusion breaks this degeneracy (since the occupied fermionic state has higher energy). 

A quantum dot weakly coupled to the fused Majorana pair can act as a sensitive probe of its charge, which can then be deduced by capacitive coupling to a microwave resonator~\cite{Grimsmo2019,Smith2020c}.  Alternatively, the quantum dot charge may be read by a tunneling measurement from a quantum point contact to the dot~\cite{Ohm2015,Steiner2020,Munk2020}, or by a charge-sensitive scanning probe (e.g. Kelvin probe microscopy or a scanning single-electron transistor). The requirement to avoid quasiparticle poisoning (which changes the electron parity) implies an important future role for such non-contact control and readout techniques.  

Although the concept of prototype computational hardware featuring a scanning probe tip may seem far-fetched today, the high quantum information density, which could be robustly encoded in a skyrmion-vortex pair array suggests that traditional nanotechnology tools may indeed prove scalable for solitonic quantum memory. We estimate that a single qubit containing four MZM-carrying pairs with $r_{sk}=$~50\,nm will occupy $\sim$\,0.5\,$\mu$m$^2$: 1000 qubits (the accepted target for a useful quantum computer) would hence fit into a 23$\times$23$\mu$m square, which is within range for a cryogenic piezo-scanner.

Gate quantum computing with skyrmion helicity-based operations necessitates progress in two complementary directions. The first concerns developing architectures that allow fast and accurate qubit readout, while preventing undesired interactions with their surroundings. Possible platforms rely on microwave techniques, especially suitable to detect helicity shifts as a function of frequency, magnetic, and electric fields. A possibility involves skyrmions coupled to coplanar waveguide resonators with a qubit-dependent frequency, while phase shifts of the reflected readout tone can measure the quantum state of the system. Magnetic resonance local techniques introduced in detail in the preceding section can detect eigenstate transitions in individual skyrmions.

Incorporating magnetic skyrmions into scalable solid-state devices with low error rates for quantum computing presents a second challenging direction. The design of quantum hardware entails adjusting the skyrmion's properties from a bottom-up approach. The development of cleaner magnetic samples and interfaces with improved crystallinity reduces the impact of damping mechanisms. Layer-by-layer growth of skyrmion-hosting perovskites is possible by molecular beam epitaxy \cite{101063} and pulsed laser deposition \cite{CHANG2010621}. Dielectrics with low loss, such as the hexagonal boron nitride (hBN) thin films with a microwave loss tangent in the mid 10$^{-6}$ range at low temperatures, provide a platform that can be effectively integrated into skyrmion qubit architectures  \cite{Wang2022}. Skyrmions are nucleated with reversible and reproducible mechanisms \cite{Schott2017,Hsu2017}. To attain positional control it is crucial to place skyrmions in confined geometries or by defect engineering \cite{Boulle2016,Ho2019,doi:10.1073/pnas.1600197113}. To address scalability, skyrmions can be fabricated in large arrays using the lithographic technology employed in microelectronics. Since they are separated at nanoscale distances, understanding and controlling skyrmion-skyrmion interaction is crucial. 

\section{Outlook}




Academic interest in fundamental and technological aspects of quantum skyrmions is rising. While we have discussed the current theoretical understanding and experimental status here, many aspects remain to be uncovered. We close by highlighting a short - and certainly non-exclusive - list of scientific targets which we believe will lead to long-term advances in this field.

The first target concerns the interactions and dynamics of quantum skyrmions. In quantum magnets, skyrmions can be considered as quantum-mechanical quasiparticles, which possess anomalous dynamics highlighting their close relation with the theory of the quantum Hall effect. Many-skyrmion systems require further investigation, as notions of entanglement, decoherence, and the existence of strongly interacting phases in analogy to the fractional quantum Hall effect need to be elucidated. The interaction of quantum skyrmions with electrons also demands further study, especially since this may potentially be responsible for the infamous non-Fermi liquid phase in metallic B20 compounds \cite{Pfleiderer2001,Pfleiderer2004,Pedrazzini2007,Ritz2013,Kirkpatrick2010, Binz2008}.

The second objective relates to the statistics of quantum skyrmions. One of Tony Skyrme's visionary conjectures for his model of baryons was the transmutations of spins and statistics: the description of fermions in terms of bosonic textures \cite{Skyrme1988}. The underlying mechanism is a Hopf term that changes the phase when two skyrmions are interchanged, leading to emergent fermionic exchange statistics. The possible realization of a corresponding Hopf term \cite{Wilczek1983,Freed2018} for two-dimensional magnetic skyrmions needs to be carefully studied. An itinerant magnet with Dirac electrons might be a potential candidate for its realization \cite{Abanov2000}: a detailed investigation of quantum skyrmions in such systems would be a worthwhile endeavor. 

The third aim will employ skyrmions to establish topological states with quantum coherence. The challenge of building a topological quantum information platform from superconductor-skyrmion hybrids has been eased by spintronics research. 
However, the creation of atomically precise heterojunctions in these materials remains an open challenge. There is also a lack of understanding of the homogeneity in the synthetic SOC induced by chiral spin texture arrays:
a precondition for high MZM mobility during future braiding operations. Another question concerns the impact of the nanoscopic skyrmion spin structure on the order parameter in a proximate superconductor. In addition to the spinless $p_x \pm ip_y$ state, non-collinear magnetism may induce other subdominant pairing symmetries, including spinful topological superconductivity as well as isotropic odd-frequency pairing. 
Distinguishing these states 
will require characterization protocols with high sensitivity to individual symmetry components of the order parameter: spin, parity, orbital/band/valley index, and frequency~\cite{Linder2019}. Recent superfluid density measurements using methods adapted from circuit quantum electrodynamics~\cite{Phan2022,Bottcher2024} may provide a signpost towards the future, especially if such techniques are adapted to offer spatial resolution. Beyond superconducting systems, one may also envisage extending or creating new topological quantum phases by coupling spin textures to magnetic superfluids or cold atoms. 

The final goal entails the manipulation of spin helicity as a quantum variable. The quantization of the low-energy degrees of freedom of skyrmions provides a clear pathway towards functionalization. Experimental realization of helicity quantization - and local control of the helicity potential via magnetoelectric coupling - will enable a completely new class of logic elements based on nanoscale spin topology. In these qubit candidates, the continuous helicity variable of a classical skyrmion is mapped to a smaller set of discrete quanta in a nanoskyrmion.  Beyond this aptitude for traditional gate-based quantum computing, the mobile particle-like nature of magnetic skyrmions renders them well-suited to newer unconventional techniques in quantum information processing.  An attractive target would be quantum neuromorphic computing, which seeks to implement reconfigurable neural networks in brain-inspired quantum hardware~\cite{10.1063/5.0020014}. 

Successful inclusion of quantum skyrmions and/or skyrmion-induced quantum phases in future technologies will require intimate integration of (i) a fundamental understanding of the quantum aspects of topological spin textures and (ii) advanced synthesis and analytical capabilities, especially new hybrid techniques that achieve quantum-limited sensitivity to one or more system parameters. It is our hope that this Colloquium will accelerate quantum skyrmionics towards delivering on its remarkable potential.



\begin{acknowledgments}
Alexander P. Petrovi\'c and Christos Panagopoulos acknowledge support from the National Research Foundation (NRF) Singapore Competitive Research Programme NRF- CRP21-2018-0001 and the Singapore Ministry of Education (MOE) Academic Research Fund Tier 3 grant MOE-MOET32023-0003. A.P.P. is also supported by the National Science Foundation ExpandQISE award no. 2228841 and LEAPS-MPS award no. 2419041. Markus Garst is supported by the Deutsche Forschungsgemeinschaft (DFG, German Research Foundation) via SPP 2137 Project-id 403030645 and Project-id 445312953. Christina Psaroudaki is an \'{E}cole Normale Sup\'{e}rieure (ENS)-Mitsubishi Heavy Industries (MHI) Chair of Quantum Information supported by MHI. Peter Fischer acknowledges support from the  U.S. Department of Energy, Office of Science, Office of Basic Energy Sciences, Materials Sciences and Engineering Division under Contract No. DE-AC02-05-CH11231 (NEMM program MSMAG).
\end{acknowledgments}
%


\bibliography{QSColl}

\end{document}